\documentclass[journal]{IEEEtran}

\usepackage{amssymb}
\usepackage{cite}
\usepackage{amsmath}
\usepackage{xcolor}
\usepackage{multirow}
\usepackage{float}
\usepackage{colortbl}
\usepackage{booktabs}
\usepackage{array}
\usepackage{mathtools, cuted}
\usepackage{makecell}
\usepackage{amsmath}%
\usepackage{enumitem}
\usepackage{dblfloatfix}
\usepackage{graphicx}

\definecolor{mycolor}{rgb}{.949,.949,.949}

\usepackage{algorithm}
\usepackage{algorithmic}

\usepackage{tikz}

\newcommand{\mycomment}[1]{}

\usepackage{hyperref} 
\hypersetup{
    colorlinks=true,       
    linkcolor=blue,        
    citecolor=blue,        
    urlcolor=blue,         
    pdfborder={0 0 0}      
}

\begin{document}
\bstctlcite{IEEEexample:BSTcontrol}

\title{RoCoF Constrained Regional Inertia Security Region: Formulation and Application}
\author{Jiahao~Liu,~\IEEEmembership{Student Member,~IEEE}, Cheng~Wang,~\IEEEmembership{Senior Member,~IEEE}, Tianshu~Bi,~\IEEEmembership{Fellow,~IEEE}
\vspace{-35pt} 
}

\maketitle

\begin{abstract}
    The regional inertia, which determines the regional rate of change of frequency (RoCoF), should be kept in a secure status in renewable-penetrated power systems. To break away from mapping the regional maximum RoCoF with regional inertia in a linearized form, this paper comprehensively studies the regional inertia security problem from formulation to applications. Firstly, the regional inertia security region (R-ISR) is defined, whose boundary is non-linear and non-convex. Then, a local linearized method is devised to calculate the global maximum of regional RoCoF. The non-convex ISR boundary is expressed by multiple simple boundaries corresponding to each local solution, which can be obtained by a simple search-based method. Finally, the convexified R-ISR constraint is formed by convex decomposition and embedded in an inertia optimal adjustment model. The results on a 3-region system show some counter-intuitive findings, such as increasing the inertia of one region may worsen its RoCoF.
\end{abstract}
\vspace{-5pt} 
\vspace{-5pt} 

\vspace{-8pt} 
\section{Introduction\label{Sec:Introd}}
\vspace{-4pt} 

\IEEEPARstart{I}NCREASING penetration of inverter-based resource (IBR) in power systems has drastically decreased the system inertia level \cite{ZZhang21}. Based on the swing equation under a major disturbance, a lower inertia level leads to a higher rate of change of frequency (RoCoF). The out-of-limit RoCoF can cause a cascading frequency drop since RoCoF-based protection trips IBRs, leading to blackouts in real-world low-inertia power systems in recent years \cite{EventGB19}. Therefore, limiting the RoCoF value by maintaining the system inertia level is of great importance.

In low-inertia power systems, the inertia level is time-varying due to the application of the virtual inertia (VI) from IBR \cite{BTan22}. This raises the requirement for fast assessment of inertia security under the constraint of RoCoF value under major disturbances \cite{FFRNerc20}. At the same time, thanks to rapidly adjustable inertia resources like VI and the synchronous generator that can start quickly to mitigate system inertia shortages, the inertia level can be adjusted in a fast and economical manner by executing power system dispatch optimization if the inertia security is compromised \cite{BShe24, MHermans2020}.

For the assessment of inertia security, the problem is formulated as determining whether the system's inertia is higher than the minimum value under major disturbances, which is defined as critical inertia \cite{FFRNerc20} or minimum inertia \cite{HGu18}. Considering the maximum limit of dispatchable inertia, the above single-end index is extended to the security region in \cite{GZhang23} and \cite{WZhang24}. To facilitate quantitative assessment, the mapping relationship between post-disturbance maximum RoCoF, which occurs at the initial time after disturbance, and system inertia can be easily deduced based on the swing equation in the system center of inertia (COI) frame \cite{PMAnderson90}.

For the optimal dispatch of system inertia, the above mapping relationship should be converted to a RoCoF constraint. The RoCoF expression should be linearized with respect to the inertia of all dispatchable sources, which will be the decision variable in the mixed-integer linear programming (MILP) problem. Linearization is straightforward in the COI frame because the COI swing equation is linear itself \cite{PMAnderson90}, and the system total inertia is the summation of the inertia of all generators \cite{RDoherty05}. Initially, the RoCoF-constrained unit commitment (UC) dominated by synchronous generators is proposed in \cite{RDoherty05} to minimize system operation costs while ensuring the RoCoF limitation. Recently, the concept of real-time economic dispatch (ED) of virtual inertia (VI) has been established to leverage the fast response of inverter-based resources (IBR) \cite{BShe24}. Various types of VI can be included in the UC or ED problem, such as VI from battery storage \cite{CMatamala24, DPandit24, YShen23}, wind turbines \cite{CJFerrandon22, SWogrin20, ZChu20}, and HVDC \cite{YWen18}.

The inertia security based on COI frequency dynamics is well addressed by the above studies. However, due to the spatially uneven distribution of non-inertia IBR, power system frequency exhibits region-level differences, especially during the several seconds after a disturbance \cite{SYou18}; that is, the frequency of buses in different regions varies \cite{JWen21}. Ensuring COI frequency stability alone may still lead to frequency collapse because the RoCoF of low-inertia regions may swing apart from the system average dynamics and exceed the secure limitation \cite{LBadesa21pt1}. Thus, region-level inertia security should be focused on.

Analogous to studies in the system COI frame, the mapping relationship between regional maximum RoCoF and the regional inertia of all regions should be given in a linearized form. Then, the mapping relationship with the inertia of all dispatchable resources in regions can be established, and a linearized constraint can be formed. The model basis for this approach is the swing equations in the regional COI frame considering inter-region coupling \cite{LBadesa21pt1, LBadesa21pt2, PRabbanifar20, JLiu23, HGu20Zonal, MTuo22}. The time-domain solution of the regional RoCoF dynamics can be obtained.

However, unlike the system-level maximum RoCoF occurring at the initial time after disturbance, the regional maximum RoCoF occurs after several swings following the disturbance \cite{HPulgar18}. The enclosed expression of regional maximum RoCoF cannot be deduced based on the time-domain solution of regional RoCoF due to complex dynamics. Existing studies all adopt an \textbf{\textit{approximated}} approach: (i) In \cite{LBadesa21pt1} and \cite{LBadesa21pt2}, the RoCoF solution is decomposed into one system COI component and several regional COI oscillation components, which are damped trigonometric functions. The enclosed expression of the maximum value of the latter is estimated by the sum of the amplitudes of all trigonometric terms, ignoring the phase angle difference between trigonometric terms and the damping effect, giving a \textbf{\textit{conservative}} approximation. Then, the estimated expression is linearized by numerical fitting. (ii) In \cite{PRabbanifar20}, the maximum value of regional RoCoF is approximated by its initial value after the disturbance, which is mainly determined by the inertia in the same region. The subsequent steps are the same as the studies in the system COI frame. Similar approaches can be found in \cite{JLiu23} and \cite{HGu20Zonal}. (iii) In \cite{MTuo22}, the time-domain solution of regional RoCoF is fixed at certain moments to approximate its maximum value. A piecewise linearized algorithm is adopted to form the RoCoF constraint.

The conservative dispatch plan given by approach (i) will be excessively expensive, and those given by approaches (ii) and (iii) are questionable due to the highly inaccurate approximation of maximum RoCoF. More importantly, attempting to map a linearized relationship between maximum RoCoF and inertia at the regional scale is inaccurate. This mapping should be highly non-linear due to the relative swings between each pair of regional COIs \cite{HPulgar18}.

In this paper, the RoCoF constrained regional inertia security region (R-ISR) is studied comprehensively. To the best of the authors' knowledge, this is the first work to provide an accurate description of the multi-area inertia boundary considering regional RoCoF constraint. Notably, other indices used to describe frequency stability, such as frequency nadir, are not included because the region-level differences in frequency dynamics are damped over these time frames \cite{HPulgar18}. Compared with existing works, the salient features of this work are threefold.

(1) The formulation of R-ISR is provided, whose boundary accurately depicts the nonlinear relationship between regional maximum RoCoF and regional inertia of all regions in the system. The complex characteristics of regional inertia security are fully preserved, i.e., increasing the inertia in one region may worsen its RoCoF, which would be demonstrated by simulation results.

(2) Leveraging the multiple trigonometric components in RoCoF dynamics, the global maximum RoCoF is calculated using a locally linearized approach in a fast and accurate manner. Thanks to the form of the maximum RoCoF expression, the non-convex R-ISR boundary can be expressed as the union of all security boundaries built by local maximum RoCoF, which is simple and determined by a search-based approach.

(3) The applications for regional inertia security assessment and fast optimal adjustment are performed. The R-ISR is modeled in a convex and linearized manner based on convex decomposition, and the MILP model for optimally adjusting the regional inertia considering VI is established. Results show an improved economy over the existing conservative method.

\vspace{-8pt} 
\section{Definition of R-ISR\label{Sec:ProbForm}}
\vspace{-4pt} 

\vspace{-0pt} 
\subsection{Secure Inertia from Holistic System RoCoF Perspective\label{SubSec:ProbFormCOI}}
\vspace{-2pt} 

\begin{figure}[!t]
	\centering
	\includegraphics[width=0.45\textwidth]{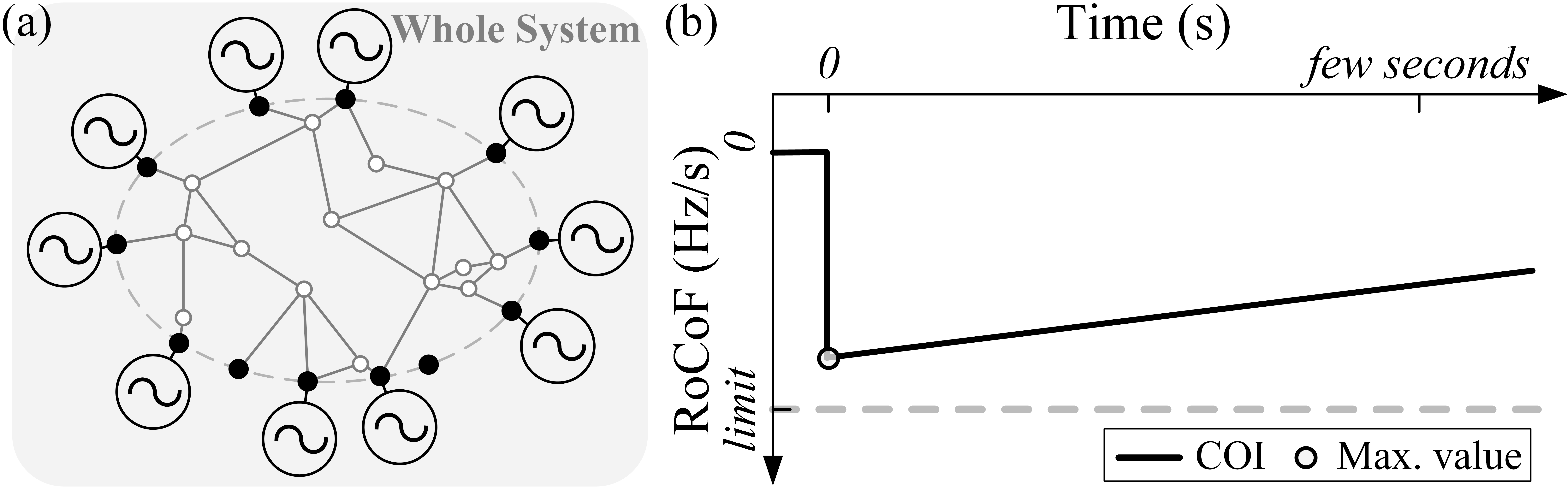}
    \vspace{-10pt} 
	\caption{(a) Topology and (b) RoCoF dynamics of a traditional power system.}
    \vspace{-10pt} 
	\label{fig:IlluSyst}
\end{figure}

In traditional power systems, inertia is distributed homogeneously on the spatial scale. As illustrated in Fig. \ref{fig:IlluSyst}(a), the system can be viewed as a whole. In this scenario, frequency dynamics can be described in the COI frame. An illustrative example is shown in Fig. \ref{fig:IlluSyst}(b), where a disturbance occurs at $t=0$, and the covered time spans a few seconds, representing the time scale of the power system inertial response \cite{PMAnderson03}. The maximum RoCoF occurs immediately after the disturbance.

From the perspective of RoCoF limitation, the secure inertia should ensure the post-disturbance maximum RoCoF does not exceed its limit \cite{FFRNerc20}. Analytically, inertia security can be derived based on the swing equation, as follows:

\vspace{-5pt} 
\begin{equation}\label{eq:SystSecu}
	H^{SYS} \geq \frac{\Delta P^{D}}{2 RoCoF^{LIM}},
    \vspace{-5pt} 
\end{equation}
where $H^{SYS}$ is the system total inertia; $\Delta P^{D}$ is the size of the disturbance; and $RoCoF^{LIM}$ is the RoCoF limit required by the local grid codes.

\vspace{-14pt} 
\subsection{Definition of R-ISR\label{SubSec:DefiIner}}
\vspace{-2pt} 

\begin{figure}[!t]
	\centering
	\includegraphics[width=0.45\textwidth]{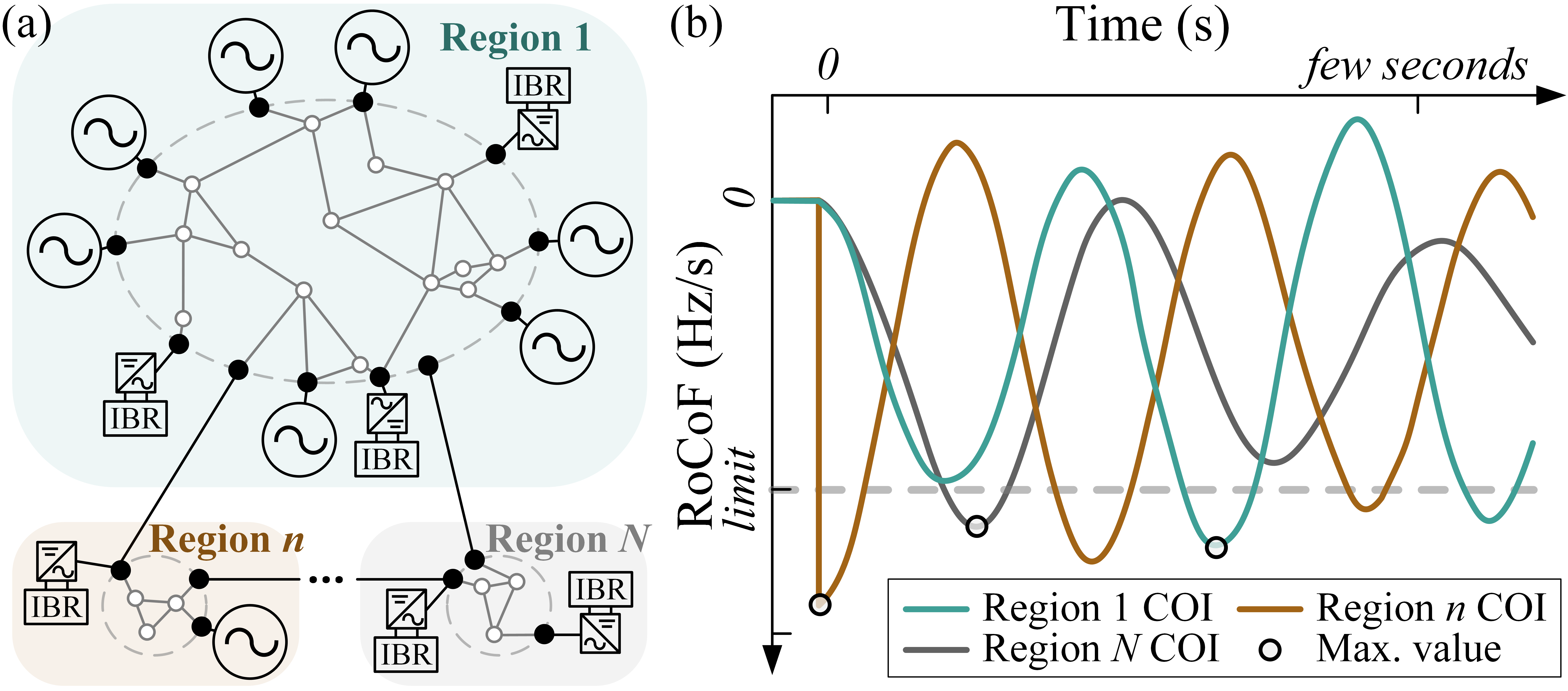}
    \vspace{-10pt} 
	\caption{(a) Topology and (b) RoCoF dynamics of an IBR-penetrated system.}
    \vspace{-28pt} 
	\label{fig:IlluArea}
\end{figure}

The increasing penetration of IBR introduces a heterogeneous distribution of inertia on the system's spatial scale. As shown in Fig. \ref{fig:IlluArea}(a), the system can be viewed as a multi-region scenario, where regions are typically partitioned based on long tie-lines connecting them \cite{YWang25} or multiple market zones \cite{MFresia24}. Each region has its own regional COI.

The post-disturbance RoCoF dynamics when a disturbance occurs in region $n$ are shown in Fig. \ref{fig:IlluArea}(b). Besides the constant decrease over the time scale of a few seconds, the oscillation introduced by the relative swinging between regional COIs can be observed. The MSN\footnote{The maximum swing number (MSN) refers to the number of swings where the maximum RoCoF occurs.} for regions $1$, $n$, and $N$ are $1$, $1$, and $2$, respectively. It can be observed that the maximum RoCoF may not occur at $t=0$. This is evident, as the RoCoF at $t=0$ for a non-disturbed region should be nearly zero.

Analogous to Section \ref{SubSec:ProbFormCOI}, the regional inertia security can be considered as the regional inertia should ensure that the regional COI RoCoF does not exceed the specified limitation. Due to the complex dynamics of a multi-region power system, an enclosed expression of regional maximum RoCoF like (\ref{eq:SystSecu}) cannot be given \cite{HPulgar18}. For a system with $N$ regions, an implicit expression of secure inertia of region $n$ can be introduced as
\vspace{-5pt} 
\begin{equation}\label{eq:RegiSecu}
	\boldsymbol{\Omega} = \left\{ \boldsymbol{H} \ \left| \ \mathcal{F} \left[ \boldsymbol{H}, \boldsymbol{x}, \Delta P^{D} \right] \leq RoCoF^{LIM} \right. \right\},
    \vspace{-5pt} 
\end{equation}
where the $\boldsymbol{H} \in \mathbb{R}_{N} $ represents the set of regional inertia of all $N$ regions; $\boldsymbol{x}$ represents the operation status of the multi-region power system, e.g., including the power transmitted between regions and the regional average voltage; $\mathcal{F} [\cdot]$ is an implicit operator for calculating regional maximum COI RoCoF. Additionally, the range limitation of regional inertia considering dispatchable resources should be expressed as
\vspace{-5pt} 
\begin{equation}\label{eq:LoUpBoun}
	\boldsymbol{\Upsilon}= \left\{ \boldsymbol{H} \ \left| \ \boldsymbol{H}^{LO} \leq \boldsymbol{H} \leq \boldsymbol{H}^{UP} \right. \right\},
    \vspace{-5pt} 
\end{equation}
where $\boldsymbol{H}^{LO} \in \mathbb{R}_{N}$ and $\boldsymbol{H}^{UP}\in \mathbb{R}_{N}$ are the lower and upper bounds of regional inertia. Then (\ref{eq:LoUpBoun}) and (\ref{eq:RegiSecu}) are combined as
\vspace{-5pt} 
\begin{equation}\label{eq:RegiSecuF}
    \boldsymbol{\tilde{\Omega}}  = \boldsymbol{\Omega} \cap \boldsymbol{\Upsilon} \! =\! \left\{ \boldsymbol{H}  \left|   
        \begin{array}{c}
            \mathcal{F} \left[ \boldsymbol{H}, \boldsymbol{x}, \Delta P^{D} \right] \leq RoCoF^{LIM} \\
            \boldsymbol{H}^{LO} \leq \boldsymbol{H} \leq \boldsymbol{H}^{UP}
            \end{array} \! \! 
        \right.
        \right\}\! .
    \vspace{-5pt} 
\end{equation}

At this time, it can be seen that the above (\ref{eq:RegiSecu})-(\ref{eq:RegiSecuF}) are in the region form. Thus, the definition of R-ISR can be given as \textbf{the set of regional inertia ensuring that the regional COI RoCoF remains within the maximum limitation}, as described by (\ref{eq:RegiSecu}). Furthermore, the R-ISR can be augmented to an enclosed region by adding the range limitation, as given in (\ref{eq:RegiSecuF}). The assessment principle of inertia security and the inertia security constraint in optimal dispatch should be: \textbf{the regional inertia is within the enclosed R-ISR}. The analytical expression is
\vspace{-5pt} 
\begin{equation}\label{eq:PricSecu}
	\boldsymbol{H} \in \boldsymbol{\tilde{\Omega}}.
    \vspace{-5pt} 
\end{equation}

\begin{figure}[!t]
	\centering
	\includegraphics[width=0.45\textwidth]{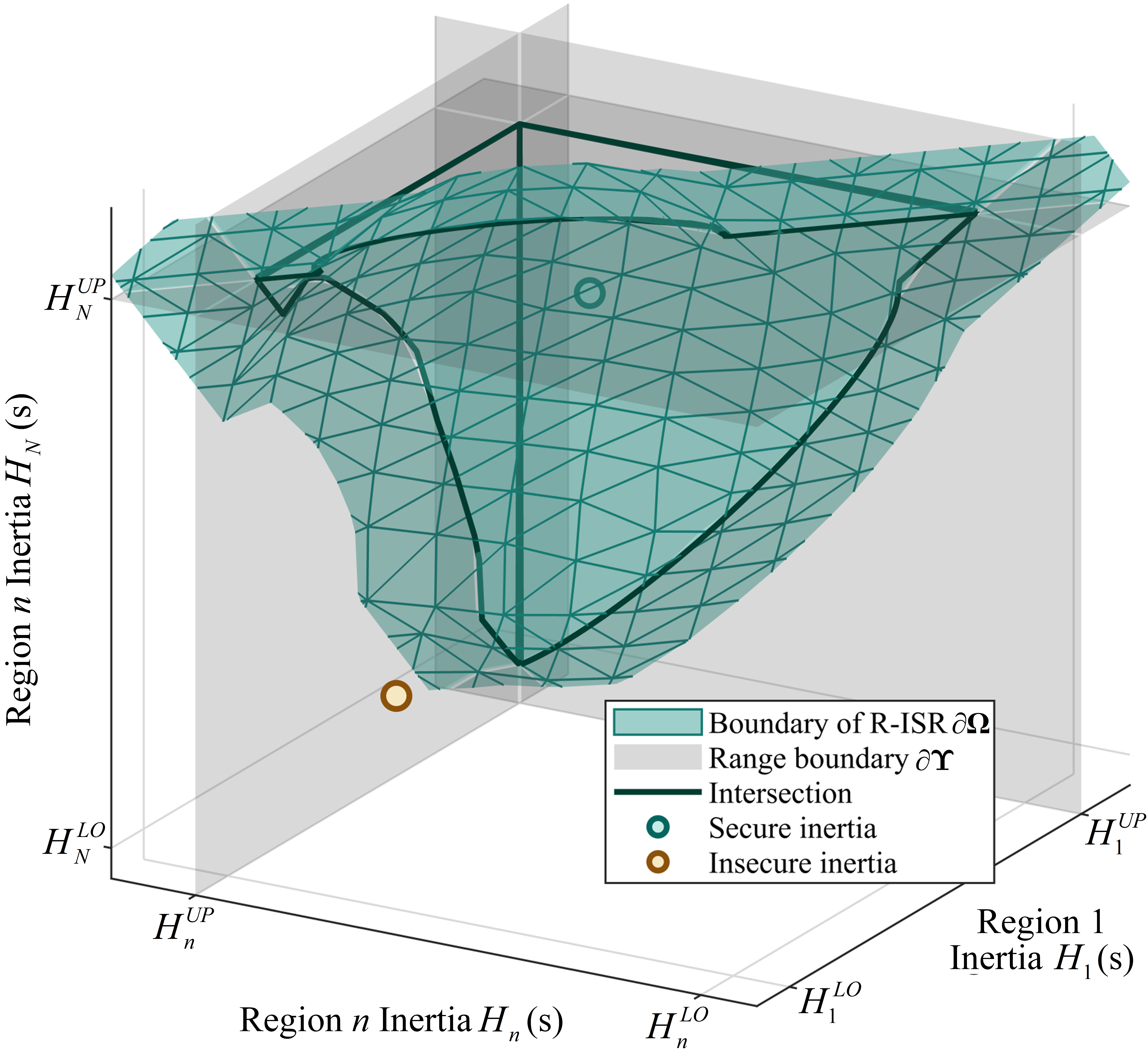}
    \vspace{-10pt} 
	\caption{Enclosed R-ISR and its boundaries.}
    \vspace{-15pt}  
	\label{fig:DefiScur}
\end{figure}

An illustrative example containing $N$ regions is shown in Fig. \ref{fig:DefiScur}. In Fig. \ref{fig:DefiScur}, $H_{1}$, $H_{n}$, and $H_{N}$ are the $1$st, $n$-th, and $N$-th elements of the set $\boldsymbol{H}$ and represent the inertia of regions $1$, $n$, and $N$. The boundary $\partial \boldsymbol{\Omega}$ of R-ISR and the boundary $\partial \boldsymbol{\Upsilon}$ of the range bound are shown separately. The intersections of these boundaries are marked, which constitute the enclosed R-ISR $\boldsymbol{\tilde{\Omega}}$. The locations of secure inertia and insecure inertia are within and outside R-ISR, respectively.

\textbf{In summary}, unlike existing studies that find a linearized mapping between regional RoCoF and regional inertia by highly approximating the regional maximum RoCoF \cite{LBadesa21pt1, LBadesa21pt2, PRabbanifar20, JLiu23, HGu20Zonal, MTuo22}, this work describes regional inertia security in the concept of R-ISR, which accurately reflects the non-linear mapping between RoCoF and inertia. This can be corroborated in Fig. \ref{fig:DefiScur}, where the shape of boundary $\partial \boldsymbol{\Omega}$ is not a plane. More importantly, since the MSN is not fixed under different inertia values, the value of the maximum RoCoF will bifurcate into a different manner when the MSN jumps. This leads to the boundary $\partial \boldsymbol{\Omega}$ being non-convex, as seen from the boundary along the $H_1$-$H_N$ plane in Fig. \ref{fig:DefiScur}. An anti-empirical conclusion that arises from this non-convexity is that increasing regional inertia can worsen the its RoCoF.

\textbf{Remark 1}: Other long-term indices, such as frequency nadir, are not included because region-level differences in frequency dynamics are significantly damped by the time these indices occur \cite{HPulgar18}. Methods for determining the nadir or other constraints at the system COI scale are well addressed in existing papers \cite{RDoherty05, BShe24}. These constraints can be incorporated with the R-ISR to develop a generalized security solution.

\textbf{Remark 2}: Consistent with existing works that focus solely on RoCoF security \cite{MTuo23}, the dynamics of primary frequency control are simplified considering the presence of a deadband in primary frequency control. Deadband limits its contribution to the maximum RoCoF occurring shortly after a disturbance.

\vspace{-14pt} 
\subsection{Challenges of Forming and Applying R-ISR\label{SubSec:ChalOnli}}
\vspace{-2pt} 

\begin{itemize}
    \item \textbf{Accurate and fast calculation of maximum RoCoF}: As a prerequisite for building R-ISR, a method for quickly calculating the post-disturbance maximum RoCoF should be established. While calculations based on time-domain simulations are accurate, they are not suitable for fast applications. The analytical approaches that are conservative or highly approximated, as given in \cite{LBadesa21pt1}, \cite{LBadesa21pt2}, \cite{PRabbanifar20}, \cite{JLiu23}, \cite{HGu20Zonal}, and \cite{MTuo22}, are inaccurate. An accurate analytical method is challenging due to complex multi-regional frequency dynamics.
    \item \textbf{Formation and convexification of R-ISR}: After establishing the method for calculating maximum RoCoF, the R-ISR $\boldsymbol{\Omega}$ should be built, primarily referring to the RoCoF-constrained boundary $\partial \boldsymbol{\Omega}$. When some initial data points on the boundary $\partial \boldsymbol{\Omega}$ are guessed roughly, the full picture of the boundary $\partial \boldsymbol{\Omega}$ can be obtained by iteratively searching for new data points along the tangent direction of the original data points \cite{TJiang21}. However, the non-convexity of the boundary $\partial \boldsymbol{\Omega}$ makes searching fail at the MSN bifurcation point. Furthermore, the optimization model with non-convex constraints is unsolvable.
\end{itemize}

\vspace{-8pt} 
\section{R-ISR Boundary Calculation \label{Sec:AnalBoun}}
\vspace{-4pt} 

Sections \ref{SubSec:StatSolu} and \ref{SubSec:MaxCalc} address the first challenge through a local linearized solving approach after fully exploring the regional RoCoF dynamics. Leveraging the form of the maximum RoCoF solution, the first part of the second challenge is addressed in Section \ref{SubSec:SearBoun} by decoupling the global non-convex boundary into several simple boundaries.

\vspace{-14pt} 
\subsection{State-Space Solution of Regional RoCoF Dynamics\label{SubSec:StatSolu}}
\vspace{-2pt} 

\begin{figure*}[!t]%
\vspace{-5pt} 
	\centering
	\begin{equation}\label{eq:LineRoco}\tag{11}
    \begin{aligned}
        \widehat {RoCoF}_{m,l} = & \Biggl( \! A_{n_{1},n_{2}}^{E} e^{\tau^{E} \hat{t}_{m,l}} \! \biggl( \! 1 \!  + \!  \tau^{E}  \left(t \! - \! \hat{t}_{m,l}\right) \! + \! \frac{{\tau^{E}}^2}{2}   \left(t \! - \! \hat{t}_{m,l}\right)^2 \biggr) \! \! + \! \!  \sum_{\forall m_{1}\in \mathbb{M}} A_{n_{1},n_{2},m_{1}}^{T} e^{\tau_{m_{1}}^{T} \hat{t}_{m,l}} \biggl( \cos \left(\theta_{n_{1},n_{2},m_{1}}^{T} \! + \! \omega_{m_{1}}^{T} \hat{t}_{m,l}\right) \! \! \! \! \! \! \! \! \! \! \\
        & +  \left(\tau_{m_{1}}^{T} \cos \left(\theta_{n_{1},n_{2},m_{1}}^{T} + \omega_{m_{1}}^{T} \hat{t}_{m,l}\right)  - \omega_{m_{1}}^{T} \sin \left(\theta_{n_{1},n_{2},m_{1}}^{T} + \omega_{m_{1}}^{T} \hat{t}_{m,l}\right) \right) \left(t - \hat{t}_{m,l} \right) \\
        & + \frac{1}{2} \left( \left({\tau_{m_{1}}^{T}}^2 \! - \! {\omega_{m_{1}}^{T}}^2\right) \cos \left(\theta_{n_{1},n_{2},m_{1}}^{T} \! + \! \omega_{m_{1}}^{T} \hat{t}_{m,l}\right) \! - \! 2 \tau_{m_{1}}^{T} \omega_{m_{1}}^{T}  \sin \left(\theta_{n_{1},n_{2},m_{1}}^{T} \! + \! \omega_{m_{1}}^{T} \hat{t}_{m,l}\right) \right) \left(t \! - \! \hat{t}_{m,l} \right)^2 \biggr) \! \! \Biggr) \! \frac{\Delta P_{n_{2}}^{D}}{2 H_{n_{2}}} \! \! \! \! \! \! \! \\
    \end{aligned}
    \vspace{-5pt} 
\end{equation}
	\hrule
\vspace{-20pt} 
\end{figure*}

To provide the model foundation for calculating the regional maximum RoCoF, the state-space model of regional RoCoF dynamics should be analyzed.

Assume a power system has $N$ regions, and the set of all regions is denoted as $\mathbb{N}$. The regional COI dominated state-space model in the inertia response time scale can be given as follows:
\vspace{-5pt} 
\begin{equation}\label{eq:StatSpac}
\begin{aligned}
    \textnormal{d} \! \! \left[\begin{matrix} \!
            \Delta \boldsymbol{\omega} \\
            \Delta \boldsymbol{\delta}
           \! \end{matrix}\right] \! \! / \textnormal{d}t \! = \! &
        \left[\begin{matrix} \!
            -\Lambda_{\left[ \boldsymbol{D} \right]}(2 \Lambda_{\left[ \boldsymbol{H} \right]})^{-1}  \! \!  \! \! \! \!& \boldsymbol{K} \\
            \boldsymbol{\omega}^{0} & \boldsymbol{0}
           \!\end{matrix}\right]  \! \! \! \left[\begin{matrix}
            \Delta \boldsymbol{\omega} \\
            \Delta \boldsymbol{\delta}
          \end{matrix}\right] 
         \! \!- \!  \!\left[\begin{matrix} \!
            (2 \Lambda_{\left[ \boldsymbol{H} \right]})^{-1}  \\
            \boldsymbol{0}
           \! \!\end{matrix}\right]  \! \!\Delta \boldsymbol{P}^{D} \!,
\end{aligned}
\vspace{-5pt} 
\end{equation}
where $\Delta \boldsymbol{\omega}  \in \mathbb{R}_{N} $ and $\Delta \boldsymbol{\delta} \in \mathbb{R}_{N} $ are the vectors of regional COI frequency and phase angle, respectively; $\Delta \boldsymbol{P}^{D} \in \mathbb{R}_{N} $ is the vector composed of the disturbance power of all regions, and the element can be $0$ if there is no disturbance in this region; $\boldsymbol{\omega}^{0}\in \mathbb{R}_{N\times N} $ is the diagonal matrix whose elements are the system nominal frequency; $\Lambda_{\left[ \boldsymbol{H} \right]}\in \mathbb{R}_{N\times N} $ and $\Lambda_{\left[ \boldsymbol{D} \right]}\in \mathbb{R}_{N\times N}$ are the diagonal matrices of regional inertia and damping coefficients; $\boldsymbol{K}\in \mathbb{R}_{N\times N}$ consists of the synchronized power coefficients between regions and is given in Appendix A \cite{AddDoc}.

The solution of (\ref{eq:StatSpac}) can be given based on eigenvalue analysis, as follows:
\vspace{-5pt} 
\begin{equation}\label{eq:ExprRoCo}
    	\boldsymbol{RoCoF} \! =\! \textnormal{d} \Delta  \boldsymbol{\omega} / \textnormal{d} t \! =\! \boldsymbol{V} e^{\boldsymbol{\Lambda} t} \boldsymbol{W} \left[\begin{matrix}
            -(2 \Lambda_{\left[ \boldsymbol{H} \right]})^{-1} \! \! \! \! \! \! &
            \boldsymbol{0}
          \end{matrix}\right]^{\top} \! \! \Delta \boldsymbol{P}^{D},
    \vspace{-5pt} 
\end{equation}
where $\boldsymbol{\Lambda}\in \mathbb{C}_{2N\times 2N}$ is the diagonal matrix whose element named $\lambda_{k}$ represents the $k$-th eigenvalue in the total $2N$ eigenvalue, and let all eigenvalues belong to set $\mathbb{S}$; $\boldsymbol{V}\in \mathbb{C}_{2N}$ and $\boldsymbol{W}\in \mathbb{C}_{2N}$  are composed by left and right eigenvectors, respectively.

Splitting one row in  (\ref{eq:ExprRoCo}) and considering only the disturbance at region $n_{2}$, the RoCoF of the region $n_{1}$ can be described as
\vspace{-5pt} 
\begin{equation}\label{eq:ExprRoCoi}
    	RoCoF =\sum_{\forall s \in \mathbb{S}} \left( \boldsymbol{v}_{s}[n_{1}] e^{\lambda_{s} t} \boldsymbol{w}_{s}[n_{2}] \right) \frac{\Delta P^{D}_{n_{2}}}{2 H_{n_{2}}} ,
    \vspace{-5pt} 
\end{equation}
where $\boldsymbol{v}_{s}\in \mathbb{C}_{N}$ and $\boldsymbol{w}_{s}\in \mathbb{C}_{N}$ represent the $s$-th left and right eigenvectors; and $\boldsymbol{v}_{s}[n_{1}]$ and $\boldsymbol{w}_{s}[n_{2}]$ are the $n_{1}$-th and $n_{2}$-th elements in eigenvectors.

Now, (\ref{eq:ExprRoCoi}) can be transformed into real number form. Note that the rank of $\boldsymbol{K}$ in (\ref{eq:StatSpac}) is $N-1$. The first two eigenvalues are real numbers, and the others are conjugates. The real number one corresponds to one damped exponential component and the others correspond to $N-1$ damped trigonometric components, as follows
\vspace{-5pt} 
\begin{equation}\label{eq:RoCoCos}
\begin{aligned}
    RoCoF =  & \biggl(  A_{n_{1},n_{2}}^{E} e^{\tau^{E} t} + \sum_{\forall m\in \mathbb{M}}   A_{n_{1},n_{2},m}^{T} e^{\tau_{m}^{T} t} \\
    & \times \cos\left(\omega_{m}^{T} t + \theta_{n_{1},n_{2},m}^{T}\right)\biggl) \frac{\Delta P_{n_{2}}^{D}}{2 H_{n_{2}}},
\end{aligned}
    \vspace{-5pt} 
\end{equation}
where $m$ is the index for the trigonometric component and $\mathbb{M}$ is the set of all trigonometric components. The detailed expressions of all coefficients are given in Appendix B \cite{AddDoc}.

\begin{figure}[!t]
	\centering
	\includegraphics[width=0.45\textwidth]{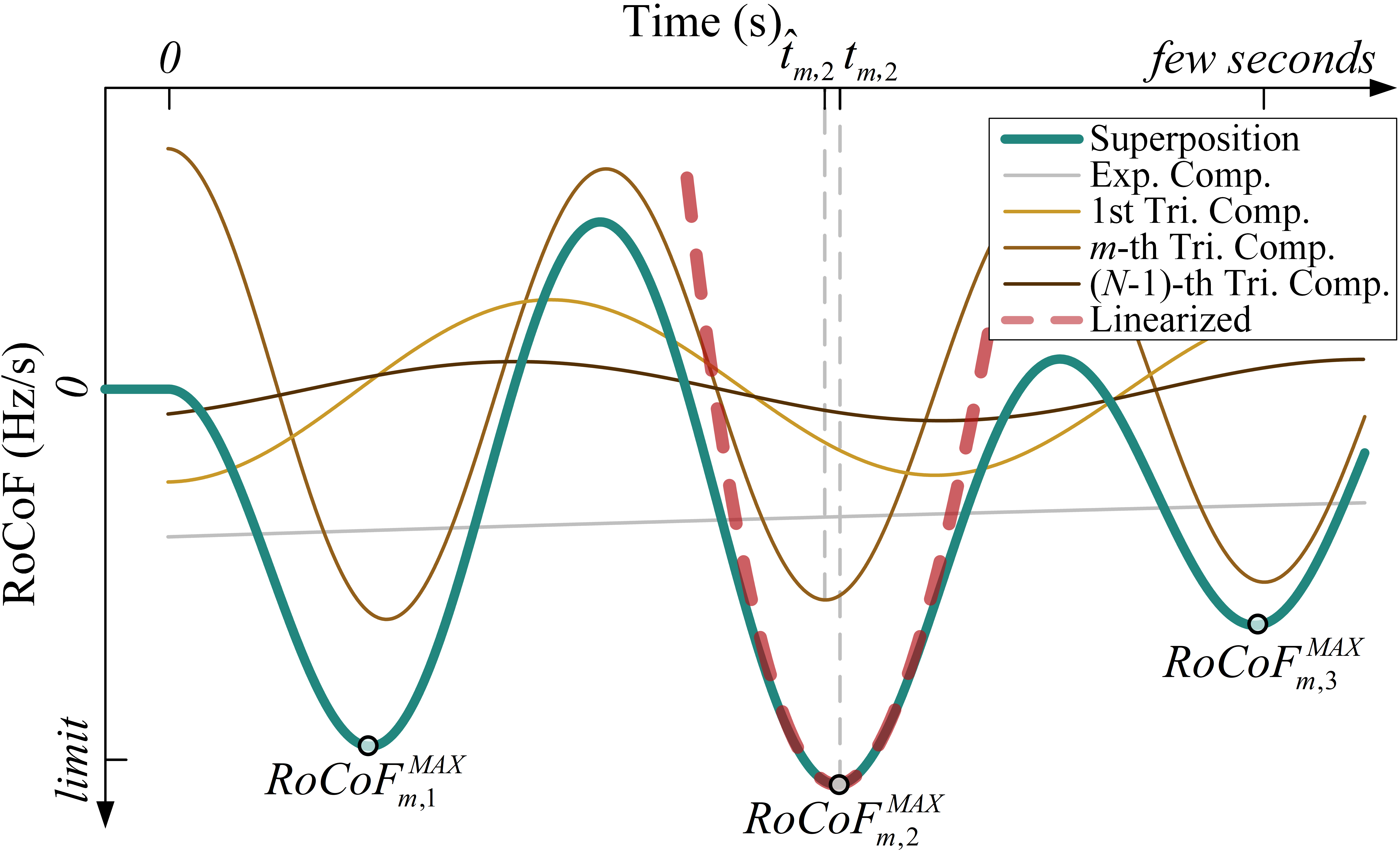}
    \vspace{-10pt} 
	\caption{Components of RoCoF dynamics and its maximum values.}
    \vspace{-20pt} 
	\label{fig:SupeRoCo}
\end{figure}

Equations (\ref{eq:RoCoCos}) show that the post-disturbance dynamics of regional RoCoF are the superposition of one damped exponential component and several damped trigonometric components, as illustrated in Fig.  \ref{fig:SupeRoCo}.

\vspace{-14pt} 
\subsection{Local Linearization Based Maximum RoCoF Calculation\label{SubSec:MaxCalc}}
\vspace{-2pt} 

The maximum value of (\ref{eq:RoCoCos}) cannot be directly solved due to its complexity. A local linearization-based approach is proposed to give an accurate analytical solution.

Seen from the superposition form of multiple trigonometric components, the possible occurrence of the maximum RoCoF is near one extrema of one trigonometric component. For example, in Fig. \ref{fig:SupeRoCo}, the maximum value occurs near the $2$nd negative extrema of the $m$-th component. Thus, after identifying all extrema of all trigonometric components, the global maximum can be determined as the maximum of these local maximums. In the following, the latter is calculated using a linearized solution approach.

First, the occurrence time of a local maximum can be estimated using the extrema of the trigonometric component. For the $l$-th extrema of the $m$-th trigonometric component in (\ref{eq:RoCoCos}), it is given as follows:
\vspace{-5pt} 
\begin{equation}\label{eq:tlin}
    \hat{t}_{m,l}=\frac{\left(2l+1\right)\pi -\theta_{n_{1},n_{2},m}^{T}}{\omega_{m}^{T}},
    \vspace{-5pt} 
\end{equation}
where $l \in \mathbb{L}$ represents the number of extrema.

Then, the local maximum value of the RoCoF expression (\ref{eq:RoCoCos}) can be solved if it is Taylor linearized to the second or third order (the root is difficult to find for orders higher than three). Taking the second order as an example, (\ref{eq:RoCoCos}) can be linearized as (\ref{eq:LineRoco}), where $m_1$ is also the index for the trigonometric component, the same as the index $m$ mentioned previously.

\setcounter{equation}{11}

The occurrence time of the local maximum of $\widehat{RoCoF}_{m,l}$ can be calculated by solving $\textnormal{d} \widehat{RoCoF}_{m,l} / \textnormal{d} t = 0$, denoted as $t_{m,l}$, and the detailed expression is shown in the Appendix C \cite{AddDoc}. The correctness of the linearized solution can be checked if the time difference between $t_{m,l}$ and $\hat{t}_{m,l}$ is small. This eliminates the case where the local extrema of the trigonometric component does not correspond to a local maximum point of the superposed RoCoF dynamics. For example, when the amplitude of one trigonometric component is small, its extrema values do not contribute to the value of the superposed RoCoF. According to the above, the expression of the local maximum RoCoF corresponding to the $l$-th extrema of the $m$-th trigonometric component can be expressed as follows:
\vspace{-5pt} 
\begin{equation}\label{eq:LocaRoco}
    RoCoF_{m,l}^{MAX} \! =\!  \left\{ \! \!  \begin{array}{ll}
    	    \nexists, \! \! \! & \left| t_{m,l} - \hat{t}_{m,l} \right| \! \geq \! \epsilon^{T},\\
    	    \widehat {RoCoF}_{m,l} \left[ t_{m,l} \right], \! \! \! &  \left| t_{m,l} - \hat{t}_{m,l} \right| \! < \! \epsilon^{T},
         \end{array}\right.\! \! \! \! \! \! \! 
    \vspace{-5pt} 
\end{equation}
where $\nexists$ indicates that the local maximum does not exist; $\epsilon^{T}$ is the threshold for the time difference.

Finally, the global maximum RoCoF can be expressed as the maximum of all local maximums, which are calculated by all extrema of all trigonometric components, as follows:
\vspace{-5pt} 
\begin{equation}\label{eq:GlobMax}
    	RoCoF^{MAX} = \max_{ \forall m \in \mathbb{M}, \forall l \in \mathbb{L} }  \left\{ RoCoF_{m,l}^{MAX} \right \}.
    \vspace{-5pt} 
\end{equation}

\vspace{-14pt} 
\subsection{Searching Based R-ISR Boundary Generation\label{SubSec:SearBoun}}
\vspace{-2pt} 

The full picture of the security boundary can be obtained by a search-based method when the shape of the boundary is convex \cite{TJiang21}. Leveraging the form of (\ref{eq:GlobMax}), the existing search-based method is used for determining the security boundary $\partial \boldsymbol{\Omega}$, which is non-convex due to bifurcation caused by MSN jumps, as mentioned in Section \ref{SubSec:ChalOnli}.

According to (\ref{eq:GlobMax}), the boundary $\partial \boldsymbol{\Omega}$ can be viewed as the union of the boundaries built by all local maximums, as follows:
\vspace{-5pt} 
\begin{equation}\label{eq:BoudMax}
    	\partial \boldsymbol{\Omega} = \underset{\forall m \in \mathbb{M}, \forall l \in \mathbb{L} }{\cap} \left\{ \partial \boldsymbol{\Omega}_{m,l}\right \},
    \vspace{-5pt} 
\end{equation}
where $\partial \boldsymbol{\Omega}$ and $\partial \boldsymbol{\Omega}_{m,l}$ are composed by global $RoCoF^{MAX} $ and local $RoCoF_{m,l}^{MAX}$ in (\ref{eq:GlobMax}), respectively.

\begin{figure}[!t]
	\centering
	\includegraphics[width=0.45\textwidth]{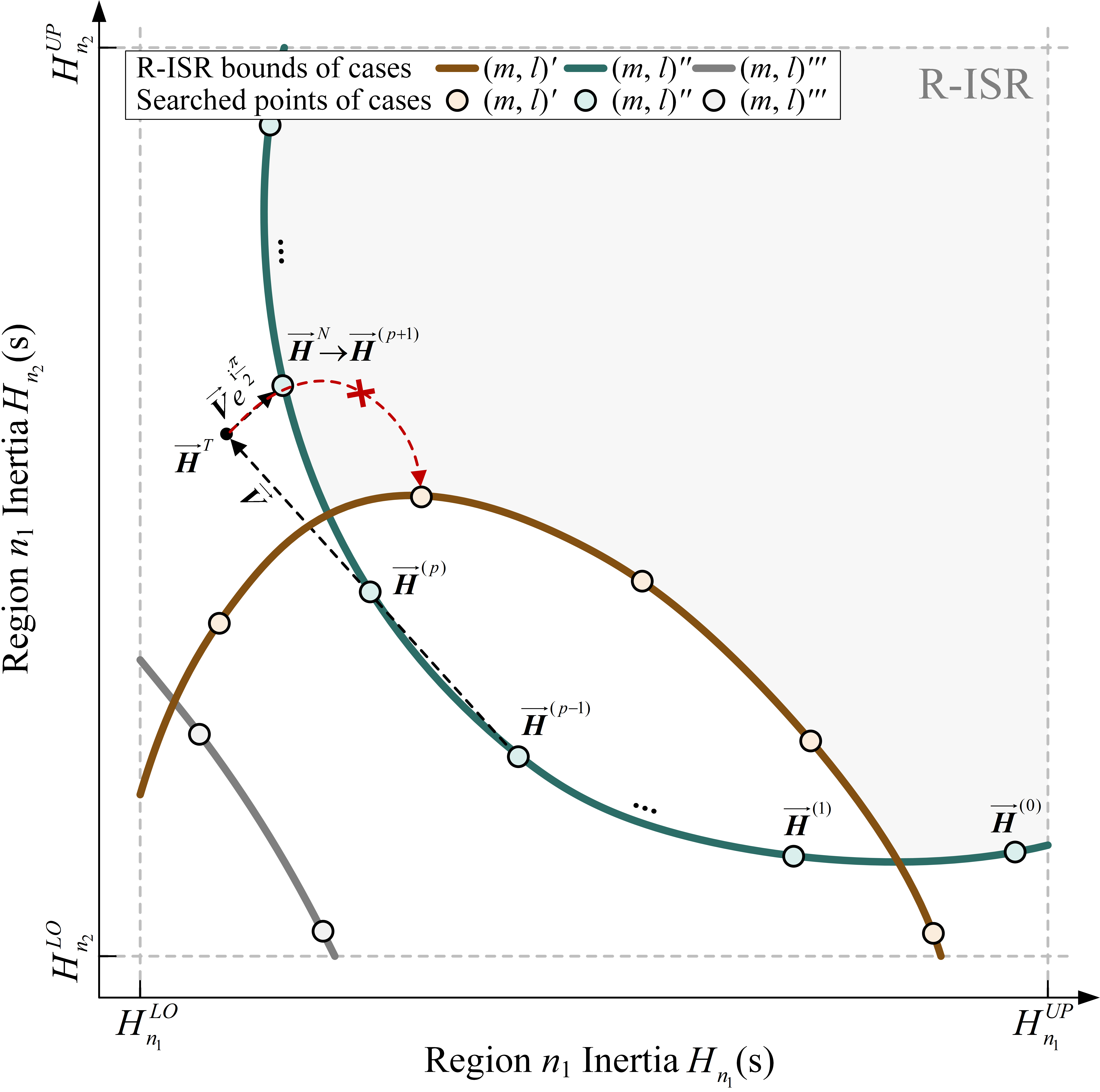}
    \vspace{-10pt} 
	\caption{Boundary of R-ISR given by searching kernel.}
    \vspace{-20pt} 
	\label{fig:InerSear}
\end{figure}

Although the full boundary $\partial \boldsymbol{\Omega}$ is non-convex, the decomposed boundary $\boldsymbol{\Omega}_{m,l}$ is relatively simple. This is because the data points on it are calculated from the same trigonometric component and the same extrema of the component, so the MSN does not change. An illustrative example in two dimensions is shown in Fig. \ref{fig:InerSear}. Three combinations of components and extrema numbers which correspond to different MSNs, $\left(m,l\right)^{\prime}$, $\left(m,l\right)^{\prime \prime}$, and $\left(m,l\right)^{\prime \prime \prime}$, are shown. The regions with adjustable inertia are $n_1$ and $n_2$.

Before discussing the full boundary, the kernel for the searching method for one component and its one extrema number is explained in \textbf{Algorithm 1}, shown in Appendix D \cite{AddDoc}. After preparing the required information, in step $1$, the vector $\vec{\boldsymbol{H}} \subset \boldsymbol{H}$ is defined to represent the values of regional inertia of two selected regions. In steps $2$ and $3$, the first two searched vectors that have a maximum RoCoF equal to the RoCoF limitation are found by random guessing. The searching process is given in Steps $4$-$14$. The vector on the security boundary in one search step is defined as $\vec{\boldsymbol{H}}^{(p)}$, which can be explained as the searched inertia whose corresponding maximum RoCoF equals the limitation at the $p$-th search step. Steps $4$ to $7$ involve searching in the tangential direction. As illustrated in Fig. \ref{fig:InerSear}, the next guess $\vec{\boldsymbol{H}}^{T}$ is given based on the tangential direction of $\vec{\boldsymbol{H}}^{(p)}$ and $\vec{\boldsymbol{H}}^{(p-1)}$ and the step size $d^T$. Steps $8$ to $13$ modify the searched inertia by searching in the normal direction. The maximum RoCoF in each search is calculated and compared with the RoCoF limitation. If the $\epsilon^{S}$ is satisfied, the searched inertia $\vec{\boldsymbol{H}}^{(p+1)}$ is considered to be the true value for this step. The exit condition is given in step $14$, where the searched inertia is beyond the upper and lower range, i.e., $\vec{\boldsymbol{H}}^{UP}=\left[H_{n_1}^{UP},H_{n_2}^{UP}\right]$ or lower bound $\vec{\boldsymbol{H}}^{LO}=\left[H_{n_1}^{LO},H_{n_2}^{LO}\right]$.

Then, the full boundary is the union of those calculated by all trigonometric components and their extrema numbers, as shown in Fig. \ref{fig:InerSear}. \textbf{Algorithm 1} can only be applied to the ``simple'' boundary, that is, the boundary calculated by the same component and extrema number. The searching process will fail if searching the entire non-convex boundary, as it cannot jump to the security boundary of another component, which corresponds to another MSN. For example, the search along the boundary of case $\left(m,l\right)^{\prime \prime}$ cannot jump to case $\left(m,l\right)^{\prime}$ after the intersection point where the MSN changes. Note that decreasing the step size of the searching kernel is useless under the non-convexity condition.

\begin{figure}[!t]
	\centering
	\includegraphics[width=0.45\textwidth]{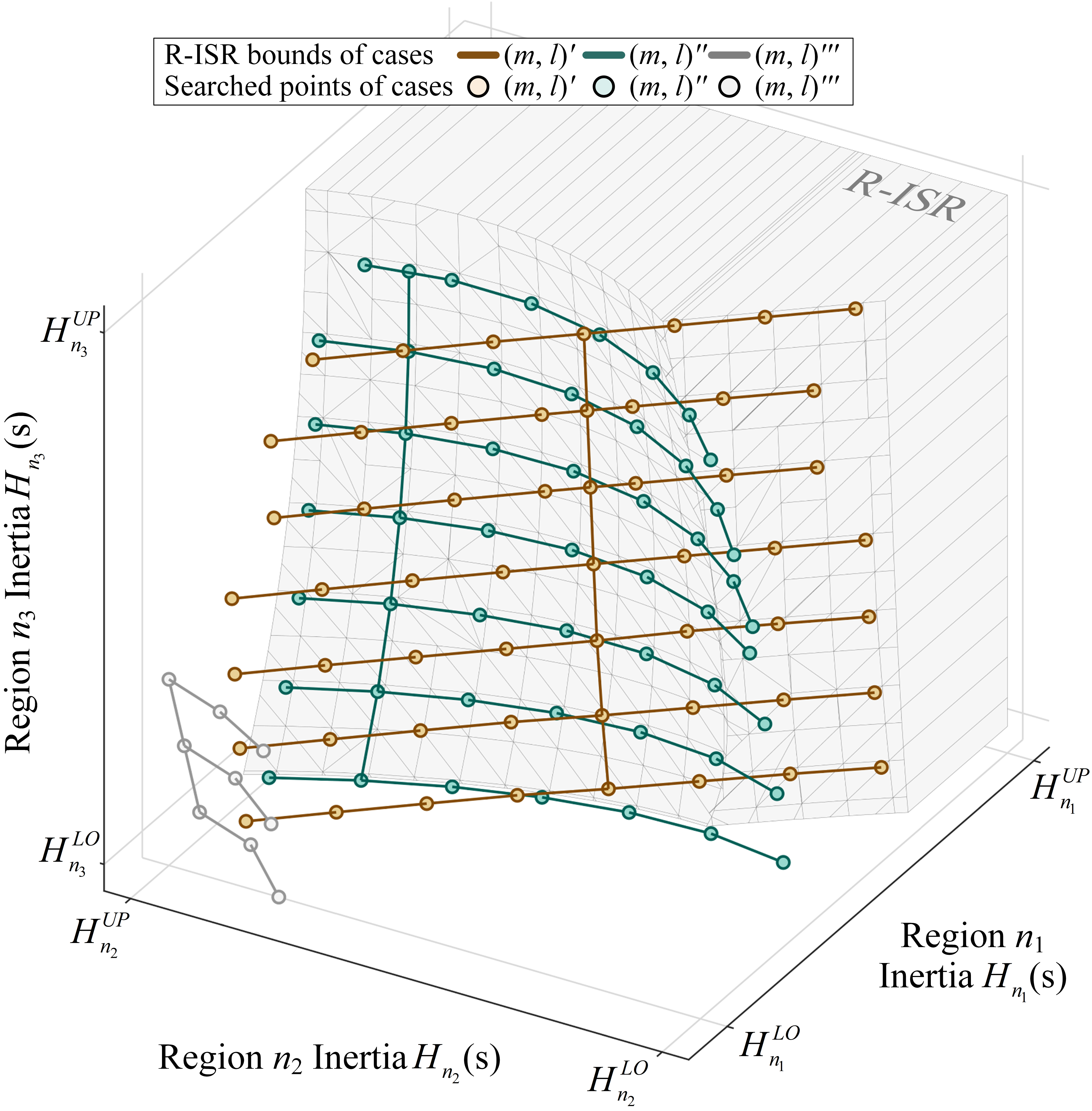}
    \vspace{-10pt} 
	\caption{Higher dimension boundary of R-ISR given by searching kernel.}
    \vspace{-20pt} 
	\label{fig:3DSear}
\end{figure}

For occasions where more than two regions need their inertia adjusted, the searching kernel can be applied to higher dimensions. The kernel should be applied for each pair of regions while the inertia of other regions remains the same. As shown in Fig. \ref{fig:3DSear}, a case involving three regions named $n_1$, $n_2$, and $n_3$ is shown. The corresponding method is explained in \textbf{Algorithm 2} in Appendix D \cite{AddDoc}. First, the inertia along $H_{n_2}$ is fixed, and the searching kernel is applied along the $H_{n_1}$-$H_{n_3}$ plane. Then, each searched point is further searched in the $H_{n_1}$-$H_{n_2}$ direction with a fixed inertia $H_{n_3}$.

\vspace{-8pt} 
\section{Convexification of R-ISR and Its Application \label{Sec:AsseOpti}}
\vspace{-4pt} 

\vspace{-0pt} 
\subsection{Convex Decomposition of R-ISR\label{SubSec:ConxDeco}}
\vspace{-2pt} 

After obtaining the searched points on the R-ISR given in Section \ref{Sec:AnalBoun}, the R-ISR can be described by a polyhedron defined by these data points, which are viewed as the vertexes of the polyhedron, as follows:
\vspace{-5pt} 
\begin{equation}\label{eq:BoudSet}
    	\boldsymbol{\tilde{\Omega}} =\left\{ \boldsymbol{H} \ \left| \ \boldsymbol{H} \in  \Gamma \! \left[  \boldsymbol{H}^{(p)}, \forall p\in \mathbb{P} \right] \right. \right\},
    \vspace{-5pt} 
\end{equation}
where $\Gamma \left[\cdot \right]$ represents the operator for defining polyhedron; $\boldsymbol{H}^{(p)}$ including the searched points and the points on the edge of inertia range limit $\boldsymbol{\Upsilon}$, and all above points are in the set $\mathbb{P}$.

As mentioned previously, the R-ISR is non-convex and cannot be used for MILP-based optimization. First, they should be decomposed into several smaller and convex parts, as follows
\vspace{-5pt} 
\begin{subequations}\label{eq:BoudCon}
\begin{equation}\label{eq:BoudCon_a}
    	\boldsymbol{\tilde{\Omega}} = \underset{\forall i}{\cup}
      \left\{  \boldsymbol{\tilde{\Omega}}^{Con}_{i}
      \right \},
    \vspace{-5pt} 
\end{equation}
    \begin{equation}\label{eq:BoudCon_exi}
    \boldsymbol{\tilde{\Omega}}_{i}^{Con} =\left\{ \boldsymbol{H} \ \left| \ \boldsymbol{H} \in  \Gamma \! \left[  \boldsymbol{H}^{(p)}, \exists p\in\mathbb{P} \right] \right. \right\} ,\quad \forall  i,
    \vspace{-5pt} 
    \end{equation}
    \begin{equation}\label{eq:BoudCon_int}
        \boldsymbol{\tilde{\Omega}}_{i}^{Con}\cap \boldsymbol{\tilde{\Omega}}_{j}^{Con} =\emptyset ,\quad \forall  i,\forall  j,
    \vspace{-5pt} 
    \end{equation}
\end{subequations}
where $i$ and $j$ are both indices for the decomposed parts. Equations (\ref{eq:BoudCon_exi}) and (\ref{eq:BoudCon_int}) are the conditions for the decomposition: the first states that the decomposed parts are defined by partial vertices of the R-ISR, and the second ensures that each two parts do not intersect.

Then, the R-ISR can be represented using the big-M trick, where auxiliary variables are introduced to represent the union operation in (\ref{eq:BoudCon_a}). Finally, condition (\ref{eq:PricSecu}) can be reformed based on (\ref{eq:BoudCon}) as follows:
\vspace{-5pt} 
\begin{subequations}\label{eq:InerCons}
    \begin{equation}\label{eq:InerCons_sum}
    \boldsymbol{H}\in   z_{i} \boldsymbol{\tilde{\Omega}}^{Con}_{i}, \quad \forall i,
    \vspace{-5pt} 
    \end{equation}
        \begin{equation}\label{eq:InerCons_z}
    \sum_{\forall i}  z_{i}=1,
    \vspace{-5pt} 
    \end{equation}
\end{subequations}
where $z_{i}$ is a binary variable, representing $z_{i} \boldsymbol{\tilde{\Omega}}_{i}^{Con}$ is $\emptyset$ when $z_{i}=0$.

Note that the above decomposition operation can be directly called by commercial software like CGAL developed in C++. This software is based on strict mathematical proofs, ensuring that no errors are introduced.

\vspace{-14pt} 
\subsection{Inertia Security Assessment and Adjustment\label{SubSec:Appl}}
\vspace{-2pt} 

The assessment of regional inertia security can be performed directly using (\ref{eq:InerCons}). If regional inertia is identified as insecure, fast re-dispatch action should be taken by starting fast-start generators or adjusting the VI settings of IBR. An MILP-based optimization model is prepared considering economic efficiency and RoCoF security. In the optimization model, the objective function includes two parts: the costs of synchronous generators (SGs) and the IBRs employing the VI control, as given in (\ref{eq:Obj}). For the costs of SGs, the first term is the operation cost, and the second and third terms represent the start-up and shut-down costs.
\vspace{-5pt} 
\begin{subequations}\label{eq:Obj}
    \begin{equation}\label{eq:Obj_min}
        \min \sum_{t} \biggl( \sum_{g} \left( C_g + \! C_g^U z_{g,t}^{U} \! + \! C_g^D z_{g,t}^{D} \right)  \! + \!  \sum_{v} \!  C_{v}^{VI} H_{v,t}^{VI}  \!\biggl) \!,
    \vspace{-5pt} 
    \end{equation}
    \begin{equation}\label{eq:Obj_C}
    C_g =C_g^{R2} {P_{g,t}^G}^2 + C_g^{R1} P_{g,t}^G + C_g^{R0},
    \vspace{-5pt} 
    \end{equation}
\end{subequations}
where $g$ and $v$ are the indices for SG and IBR equipped with VI, respectively; $t$ is the index for the time period; $P_{g,t}^{G}$, $z_{g,t}^{U}$, and $z_{g,t}^{D}$ are the decision variables representing the power output, start-up, and shut-down status of generators; $H_{v}^{VI}$ is the decision variable representing the VI settings, noting that $H_{v}^{VI}$ can be adjusted continuously; $C_g\left[\cdot\right]$ represents the cost function and $C_g^{R2}$, $C_g^{R1}$, and $C_g^{R0}$ are the coefficients of the quadratic, linear, and constant terms; $C_g^U$ and $C_g^D$ are the cost coefficients for generator start-up and shut-down; $C_{v}^{VI}$ is the cost for VI.

Considering regional dispatch criterion, the regional power should satisfy the regional power flow requirement, as follows:
\vspace{-5pt} 
\begin{equation}\label{eq:PF}
    \sum_{\forall g\in \mathbb{G}_{n}} P_{g,t}^{G} + \sum_{\forall v \in \mathbb{G}_{n}} P_{v,t}^{VI} = P_{n,t}^{A}, \quad \forall n, \forall t,
    \vspace{-5pt} 
\end{equation}
where $\mathbb{G}_{n}$ is the set including SG and IBR in region $n$; $P_{v}^{VI}$ is the power output of IBR; $P_{n}^{A}$ is the regional active power given by power flow calculation.

The regional inertia security is given in (\ref{eq:InerCons}), and it should be noted that the security region is different for each time period because the system operation status changes. The regional inertia in (\ref{eq:InerCons}) can be expressed as follows:
\vspace{-5pt} 
\begin{equation}\label{eq:Iner}
    H_{n,t}=\sum_{\forall g \in \mathbb{G}_{n}} H_{g,t}^{G} u_{g,t}^{G} + \sum_{\forall v \in \mathbb{G}_{n}} H_{v,t}^{VI} u_{v,t}^{VI}, \quad \forall n, \forall t,
    \vspace{-5pt} 
\end{equation}
where decision variables $u_{g}^{G}$ and $u_{v}^{VI}$ are the on/off status of SG and IBR; $H_{g}^{G}$ is the inertia constant of SG.

Other constraints such as generator output limits, ramping limitations, and system power flow balance should also be included, as detailed in Appendix E \cite{AddDoc}, but are omitted here for simplicity.

\vspace{-8pt} 
\section{Case Studies\label{Sec:CaseStud}}
\vspace{-4pt} 

\vspace{-0pt} 
\subsection{Test System Configuration\label{SubSec:TestSyst}}
\vspace{-2pt} 

\begin{figure}[!t]
	\centering
	\includegraphics[width=0.45\textwidth]{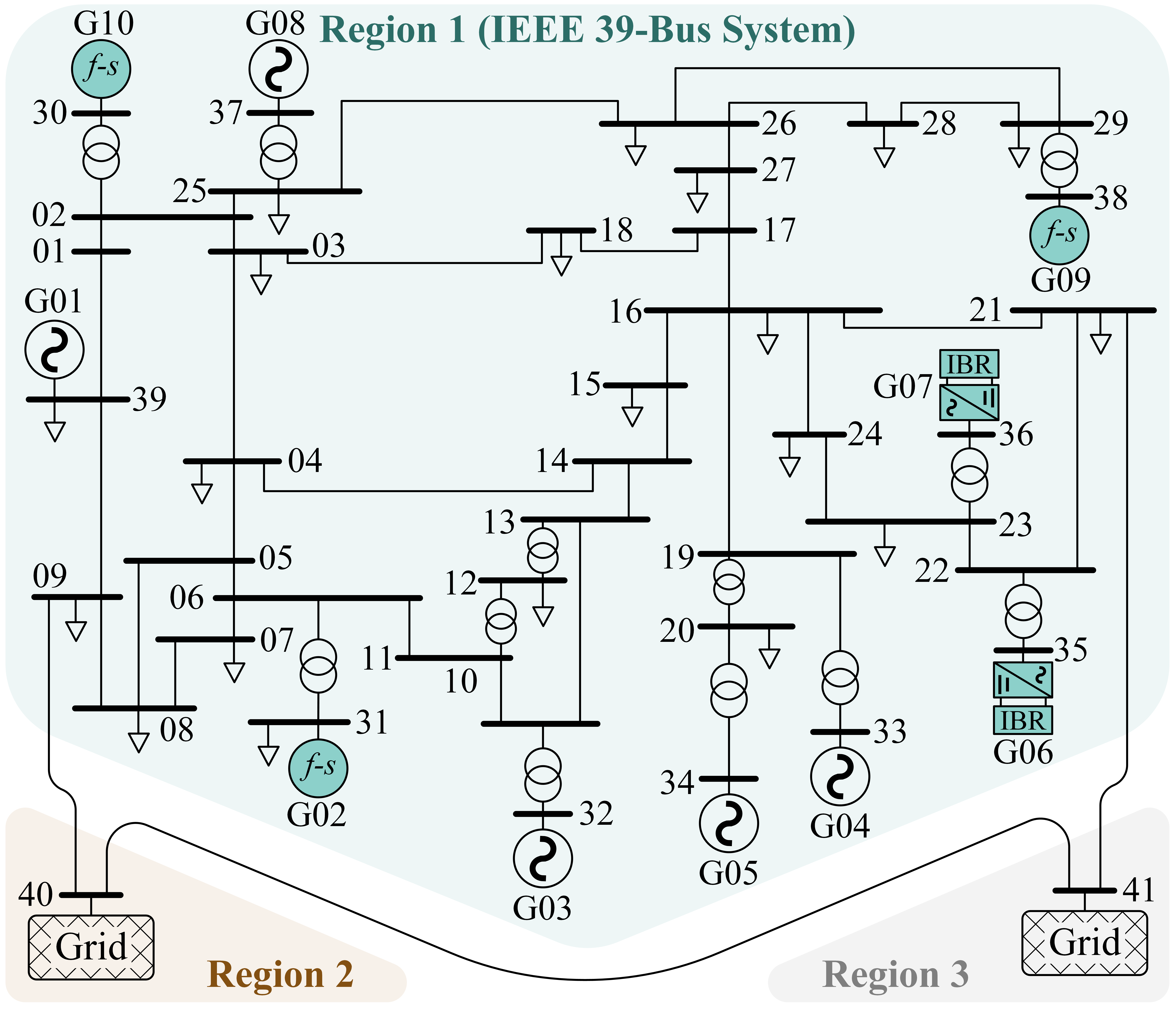}
    \vspace{-10pt} 
	\caption{Topology of 3-region test system developed from IEEE $39$-bus system.}
    \vspace{-20pt} 
	\label{fig:IEEETopo}
\end{figure}

\begin{figure}[!t]
	\centering
	\includegraphics[width=0.45\textwidth]{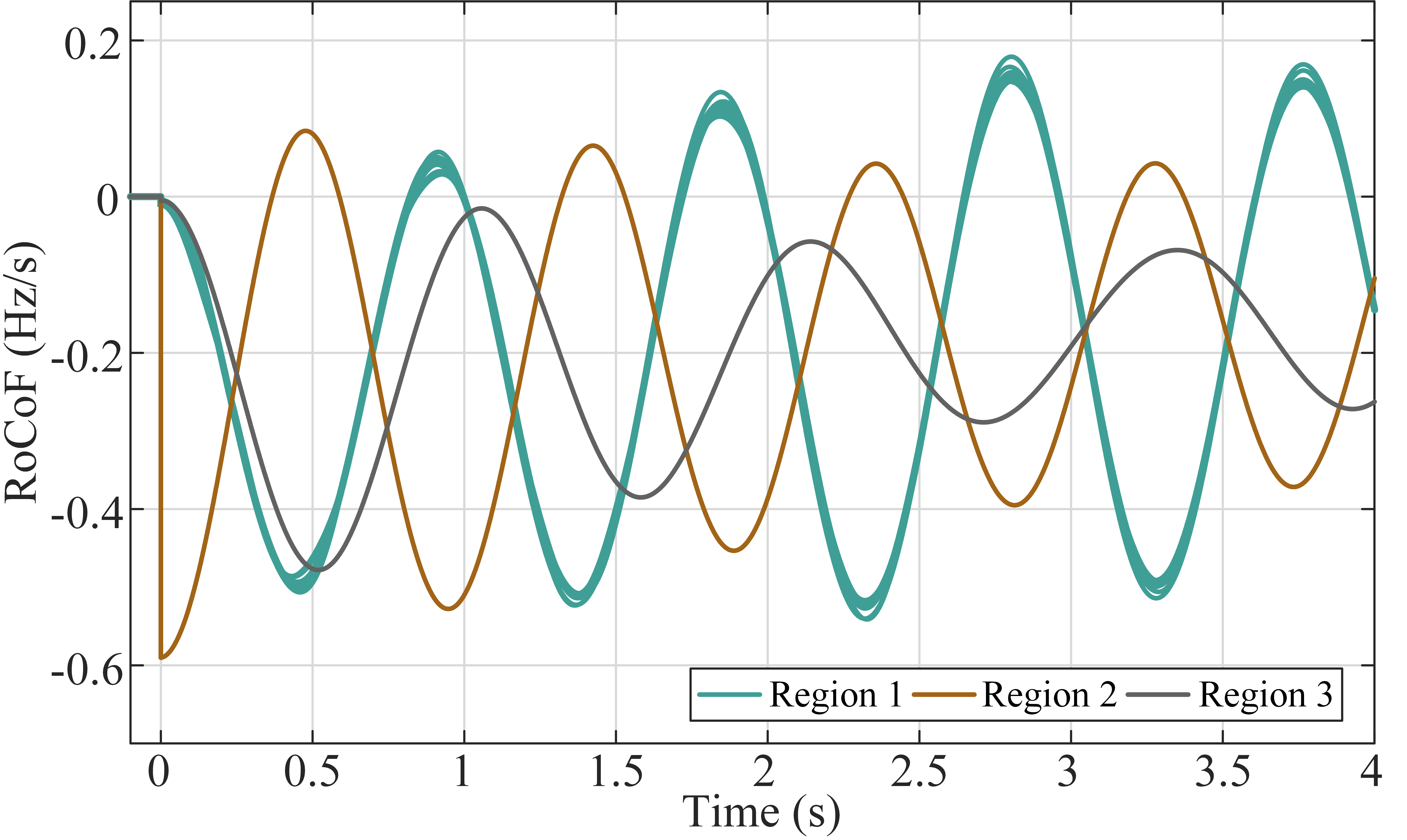}
    \vspace{-10pt} 
	\caption{Post-disturbance RoCoF dynamics.}
    \vspace{-20pt} 
	\label{fig:SystDyna}
\end{figure}

A $3$-region system is developed based on the IEEE $39$-bus system, as shown in Fig. \ref{fig:IEEETopo}. Region $1$ corresponds to the IEEE $39$-bus system, where traditional SG, fast-start SG marked in green, and IBR with VI control are included. Regions $2$ and $3$ are represented by a simplified grid composed of an equivalent generator and an equivalent load. Under certain regional inertia settings, the RoCoF dynamics under the load step located at bus $40$ are shown in Fig. \ref{fig:SystDyna}. The buses in different regions are shown in different colors. The oscillation characteristics affecting the value of maximum RoCoF and its MSN can be observed. For example, the maximum RoCoF of the region $1$ occurs at the $3$rd swing between $2$ and $2.5$ seconds. Thus, the concept of regional inertia security is important for this case.

The system parameters used in inertia dispatch are summarized in the table in \cite{AddDoc}, where the type of generator, power limits, cost coefficients, and inertia conditions are given. Note that all power-related parameters are rated at $100$ MVA. Parameter explanations classified by the type of generator are given as follows:
\begin{itemize}
    \item \textbf{Traditional SG in region $\boldsymbol{1}$}: The cost settings are common.
    \item \textbf{Fast-start SG in region $\boldsymbol{1}$}: The minimum on-off time is $0.5$ or $1$ hour \cite{MHermans2020}, much less than that of traditional ones. However, the operation costs and start-up/shut-down costs are higher than the traditional ones.
    \item \textbf{VI-equipped IBR in region $\boldsymbol{1}$}: The operation costs are smaller than those of SGs. Inertia is in a range that can be adjusted continuously.
    \item \textbf{Regions $\boldsymbol{2}$ \& $\boldsymbol{3}$}: A mixed generator is introduced for dispatch use. The operation costs are regular, and the start-up/shut-down cost are set high to avoid shutting down the entire region. The inertia includes two parts. The first refers to SG and is related to the active power level of the region, measured in seconds per unit power (i.e., s/p.u.). The second is the inertia from VI.
\end{itemize}

All studies are performed based on Intel(R) Xeon(R) Platinum 8352V CPU @ 2.10GHz and 96 GB memory. The MILP models are solved by Gurobi 11.0.1. The actual frequency dynamics is simulated by solving the power system differential-algebraic equations.

\vspace{-14pt} 
\subsection{Accuracy of R-ISR Boundary\label{SubSec:AccuBoun}}
\vspace{-2pt} 

The proposed calculation method of R-ISR boundary in Section \ref{Sec:AnalBoun} is verified. Three other methods are compared:
\subsubsection{COI-based method} The system COI RoCoF is used as a limitation for regional RoCoF. The inertia security boundary can be given based on (\ref{eq:SystSecu}) as follows:
\vspace{-5pt} 
\begin{equation}\label{eq:COIBoun}
	 \left\{ \boldsymbol{H} \ \left| \  \frac{\Delta P^{D}}{2\left(H_1 + H_2 + H_3\right) }  = RoCoF^{LIM} \right. \right\}.
    \vspace{-5pt} 
\end{equation}

\subsubsection{Conservative method} Recent approaches for addressing regional RoCoF constraints are proposed in \cite{LBadesa21pt1} and \cite{LBadesa21pt2}, claimed as the ``\textit{\textbf{conservative}}'' method. The maximum RoCoF is approximated based on (\ref{eq:RoCoCos}) without accurately calculating the extrema points. As given in (\ref{eq:ConsRoCo}), the superposition of trigonometric components in (\ref{eq:RoCoCos}) is scaled to its maximum value, as follows:
\vspace{-5pt} 
\begin{equation}\label{eq:ConsRoCo}
\begin{aligned}
    \sum_{\forall m\in \mathbb{M}} \! \! \!  A_{n_{1},n_{2},m}^{T} e^{\tau_{m}^{T} t} \cos\left(\omega_{m}^{T} t \! + \! \theta_{n_{1},n_{2},m}^{T}\right) \! \! \leq \! \! \sum_{\forall m\in \mathbb{M}} \! \! \! \!   A_{n_{1},n_{2},m}^{T},
\end{aligned} \! \! \!
    \vspace{-5pt} 
\end{equation}
where the relationship between the conservative term $\sum_{\forall m\in \mathbb{M}}   A_{n_{1},n_{2},m}^{T}$ and the regional inertia can be fitted numerically. Thus, the boundary can be expressed as follows:
\vspace{-5pt} 
\begin{equation}\label{eq:ConsBoun}
\begin{aligned}
    	\left\{ \boldsymbol{H} \ \left| \  \frac{\Delta P^{D}}{2\left(H_1 + H_2 + H_3\right) } +  \frac{m_1 H_1 + m_2 H_3}{2\left(H_1 + H_2 + H_3\right) } \right.  \right. \\
     \left. \frac{+ m_3 H_3 + m_4}{ \cdots} = RoCoF^{LIM} \right\},
\end{aligned}
    \vspace{-5pt} 
\end{equation}
where the first term represents the COI RoCoF calculated by the exponential component in (\ref{eq:RoCoCos}); $m_1$-$m_4$ are the fitted coefficients. Note that the coefficients for load level and disturbance power in \cite{LBadesa21pt1} and \cite{LBadesa21pt2} are not fitted because their impact is not considered in this section.

\subsubsection{Detailed simulation-based method} The RoCoF is generated by detailed simulation and the security boundary is built by exhaustively searching.

\begin{figure}[!t]
	\centering
	\includegraphics[width=0.45\textwidth]{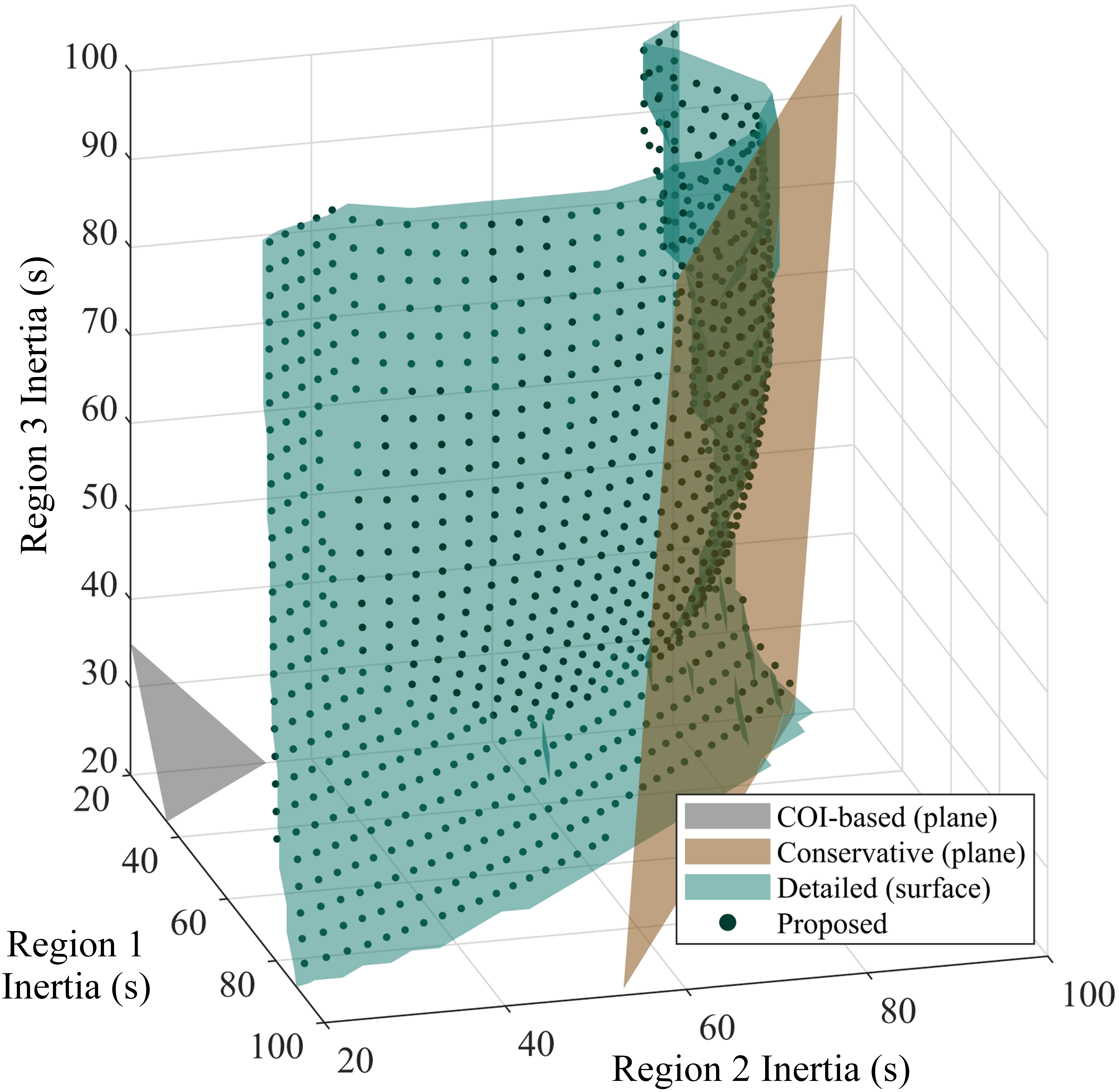}
    \vspace{-10pt} 
	\caption{Boundaries of security regions calculated by different methods.}
    \vspace{-15pt} 
	\label{fig:3DComp}
\end{figure}

The boundaries of security regions given by the proposed method and the above three methods are shown in Fig. \ref{fig:3DComp}. The RoCoF limitation is set as $0.8$ Hz/s and region $1$ is where the RoCoF is constrained. In Fig. \ref{fig:3DComp}, the boundary of the proposed method is shown by the searched data points and the other methods are given by a fitted plane or surface to provide a distinguishing view. The quantitative errors of the boundaries compared with the one calculated by detailed simulation are given in Table \ref{tab:IndeComp}. The time burden is also shown in the table.

\begin{table}[!t]
	\setlength\tabcolsep{3.0pt}
	\setlength{\aboverulesep}{0.0pt}
	\setlength{\belowrulesep}{1.0pt}
	\centering
	\caption{Quantitative indexes of different methods}
    \vspace{-8pt} 
    \begin{tabular}{ccccc}
		\toprule
		\specialrule{0em}{0.4pt}{0.4pt}
		\toprule
      Index & COI-based & Conservative & Detailed & Proposed \\
    \midrule
    Time (s) & $\approx$0 & 433   & 115200 $\approx$ 32 hours    & 600 \\
    Error (\%) & 55.78 & 24.45  & 0 & 1.12 \\
    \bottomrule
	\specialrule{0em}{0.4pt}{0.4pt}
    \bottomrule
    \end{tabular}
	\label{tab:IndeComp}
    \vspace{-20pt} 
\end{table}

Firstly, as shown in Fig. \ref{fig:3DComp}, the boundary calculated by the proposed method is close to the one calculated by detailed simulation. The error shown in Table \ref{tab:IndeComp} is only $1.12$\%. In particular, the non-convexity of the boundary is depicted accurately. For the COI-based method, the COI dynamics only capture the average dynamics, and the RoCoF overshoot introduced by swing is neglected. Thus, the calculated boundary is too optimistic, with an error of $55.78$\%, which will lead to frequency instability if used as the frequency constraint in dispatch. For the conservative method, as stated in \cite{LBadesa21pt1} and \cite{LBadesa21pt2}, the conservativeness is shown in the results compared with those calculated by detailed simulation. The error introduced is $24.45$\%. This will lead to more operating generators and higher operation costs as shown in the next subsection. Note that the boundaries of the proposed method and the conservative method overlap around $H_1=50$ s, $H_2=80$ s, and $H_3=90$ s because the equality holds in (\ref{eq:ConsRoCo}) under this inertia setting.

Regarding the time burden given in Table \ref{tab:IndeComp}, the time spent by the COI-based method is almost $0$. The time for the detailed simulation-based method is nearly $32$ hours, which is not suitable for fast application. The time burdens of the conservative method and the proposed method are almost the same and acceptable for fast use.

\vspace{-14pt} 
\subsection{Characteristics of R-ISR Boundary \label{SubSec:CharBoun}}
\vspace{-2pt} 

\begin{figure}[!t]
	\centering
	\includegraphics[width=0.45\textwidth]{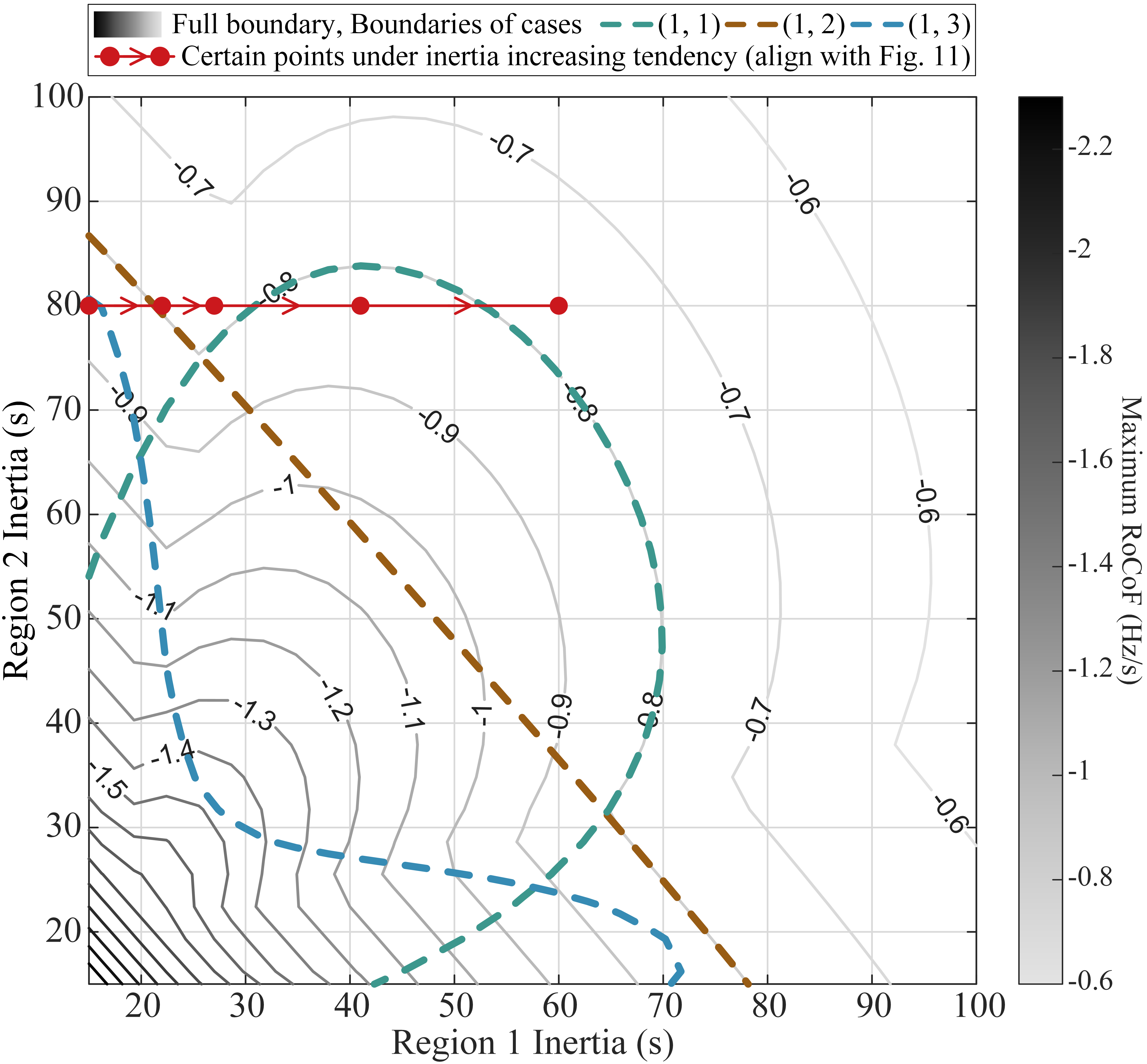} 
    \vspace{-10pt} 
	\caption{Multiple boundaries of R-ISR in contour view.}
    \vspace{-20pt} 
	\label{fig:CompAnly}
\end{figure}

The R-ISR boundary impacting the security assessment and inertia dispatch are discussed. The calculated R-ISR boundary is shown as a contour plot in Fig. \ref{fig:CompAnly}, where multiple equispaced RoCoF values are shown, from $-2.2$ to $-0.6$ Hz/s. The RoCoF of region $1$ is observed. Fig. \ref{fig:CompAnly} is given in two dimensions where only the inertia of regions $1$ and $2$ are adjusted, and the inertia of region $3$ is fixed at $90$ seconds.

\subsubsection{In the view of security assessment}
From the shape of the full boundary, as mentioned earlier, the security boundary is complex and non-convex. The COI-based or conservative method, which describes the boundary as a plane (a straight line in two dimensions), fails to assess the inertia security. Futher, it can be seen that the gradient of the contour is higher when the inertia level is low, so the regional inertia security assessment is more important under low inertia situations.

From the view of the boundary superposition, the boundary under the maximum value of $-0.8$ Hz/s is discussed further. As depicted by (\ref{eq:RoCoCos}), the $3$-region system has $2$ trigonometric components in RoCoF dynamics. The decomposed boundaries calculated under the $1$st, $2$nd, and $3$rd swing numbers of the $1$st trigonometric component are shown in Fig. \ref{fig:CompAnly} in different colors, named cases $(1, 1)$, $(1, 2)$, and $(1, 3)$, respectively. It should be noted that the boundaries calculated by the $2$nd component do not satisfy the time threshold given in (\ref{eq:LocaRoco}) and are neglected. As can be seen, the full boundary is the union of those calculated under cases $(1, 1)$ and $(1, 2)$. The boundary under case $(1, 3)$ is too loose to be considered. The shapes of the boundaries under cases $(1, 1)$ and $(1, 2)$ are totally different, being nearly linear and curved, respectively. This should be noticed in security assessment to grasp the security trend around the current operating point.

\subsubsection{In the view of inertia dispatch}
The non-convexity leads to a special circumstance in inertia dispatch in a regional system. The red arrowed line in Fig. \ref{fig:CompAnly} represents increasing the inertia of region $1$ while keeping the inertia of region $2$ fixed. Starting from inertia of $10$ seconds, the maximum RoCoF along the arrow direction is reduced to $-0.8$ Hz/s when increasing the inertia to $31$ seconds. However, the RoCoF will increase if the inertia keeps increasing to around $45$ seconds. This means \textbf{increasing the inertia in one region may worsen its post-disturbance RoCoF}.

\begin{figure}[!t]
	\centering
	\includegraphics[width=0.45\textwidth]{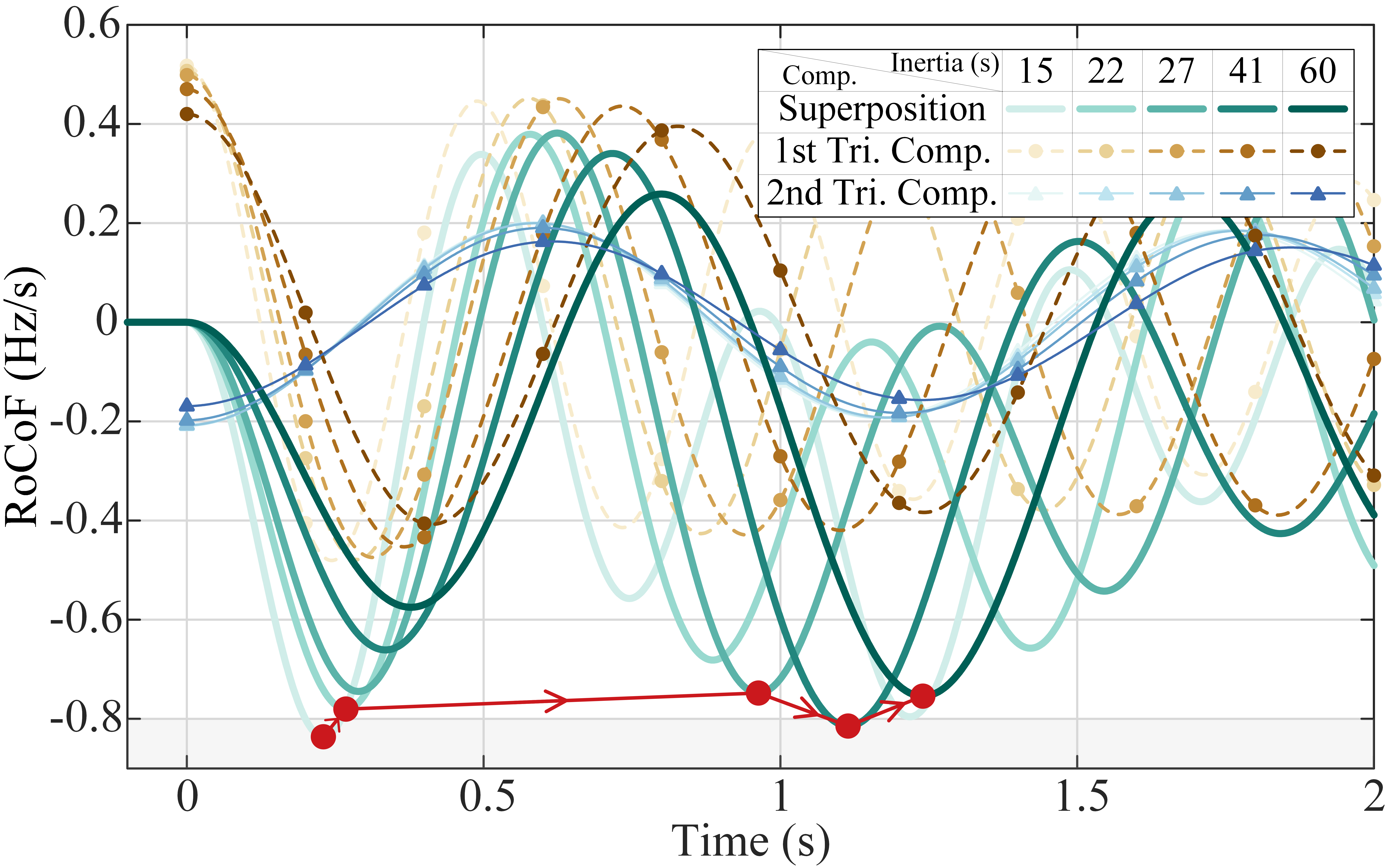} 
    \vspace{-10pt} 
	\caption{RoCoF dynamics with increasing inertia of region 1.}
    \vspace{-20pt} 
	\label{fig:FreqPlot}
\end{figure}

To observe closely, Fig. \ref{fig:FreqPlot} shows the RoCoF dynamics and its $1$st and $2$nd trigonometric components. The cases where the inertia of region $1$ equals $15$, $22$, $27$, $41$, and $60$ seconds are shown. For the inertia of $15$ or $22$ seconds, the amplitude of all components decreases when inertia increases, and the dynamics of the superposed signal are depressed, which is common sense. However, when inertia increases to $27$ seconds, the MSN jumps to the $2$nd swing since the extrema of the $1$st and $2$nd components are closer. When increased further to $41$ seconds, the closer proximity leads to the maximum value at the $2$nd MSN increasing. After that, at $60$ seconds, although the extrema points are closer and overlapped, the impact of increasing inertia on depressing the amplitude takes a leading role and the RoCoF starts to decrease again. Considering the above complex characteristics, the proposed R-ISR based inertia adjustment method is necessary.

\vspace{-14pt} 
\subsection{Regional Inertia Optimal Adjustment\label{SubSec:InerDisp}}
\vspace{-2pt} 

The optimal fast adjustment of regional inertia is verified. The dispatch interval is set at $0.5$ hour, and dispatching for the future $4$ hours is considered for fast scheduling after inertia assessment shows an unsatisfied situation. The load of region $1$ increases from hour $0.5$ to hour $2$ and decreases from hour $2.5$ to hour $4$. Regions $1$ and $2$ are dispatchable, and the inertia of region $3$ is fixed to provide a clear result in two dimensions. The region where the RoCoF is to be constrained is region $1$, with a RoCoF limitation of $-0.8$ Hz/s. The considered disturbance is the load increase at bus $40$. All generators in the table in \cite{AddDoc} are dispatchable. The COI-based method and the conservative methods are compared. The security regions can be embedded in the UC model easily since the boundary described by (\ref{eq:COIBoun}) and (\ref{eq:ConsBoun}) is in a linearized relationship with decision variables, i.e., regional inertia or generator inertia. Note that the dispatchable range of regions $1$ and $2$ is bounded between $10$-$70$ and $40$-$100$ seconds.

\begin{figure}[!t]
	\centering
	\includegraphics[width=0.45\textwidth]{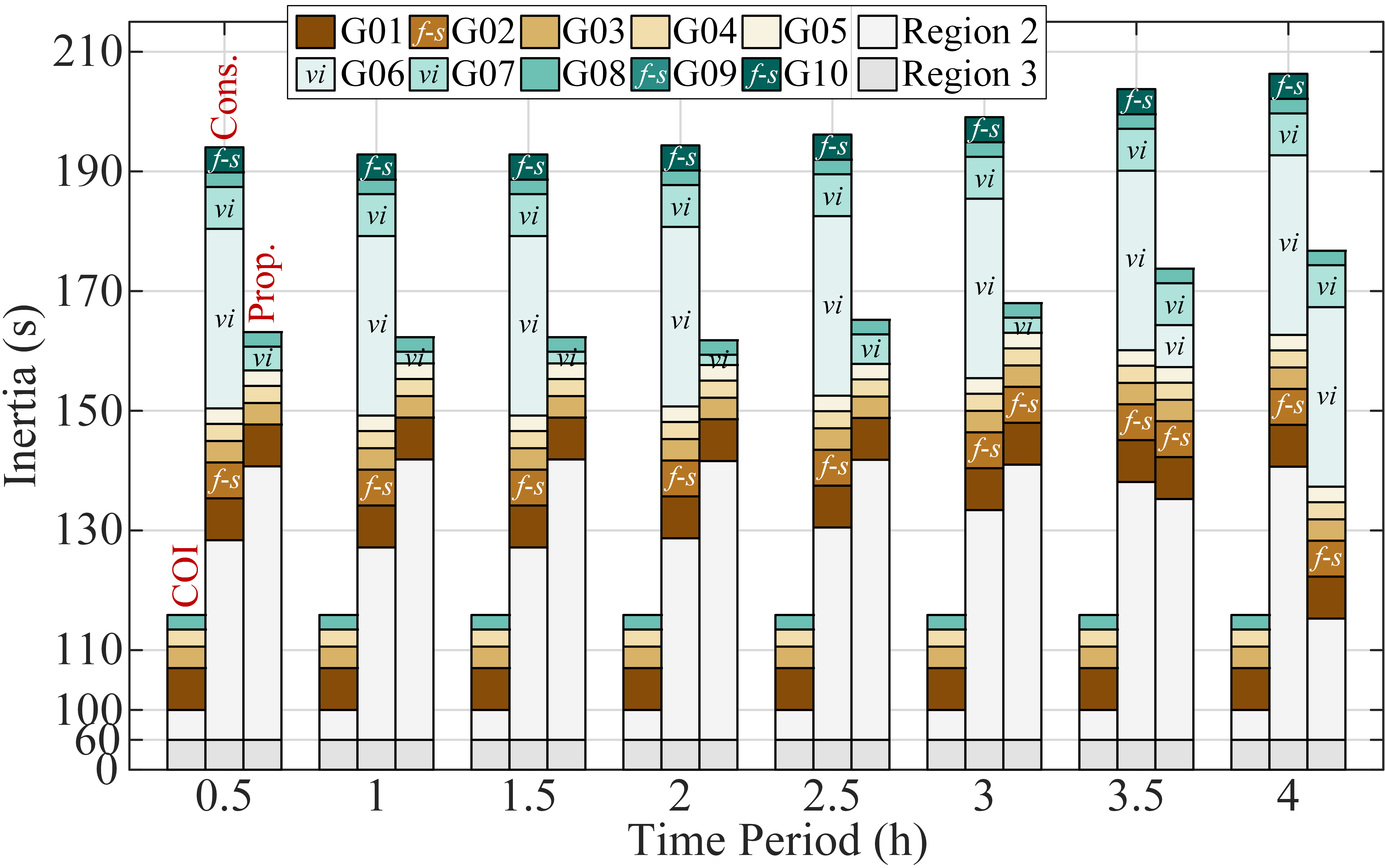} 
    \vspace{-10pt} 
	\caption{Generation scheduling results.}
    \vspace{-14pt} 
	\label{fig:UCResu}
\end{figure}

\begin{figure}[!t]
	\centering
	\includegraphics[width=0.45\textwidth]{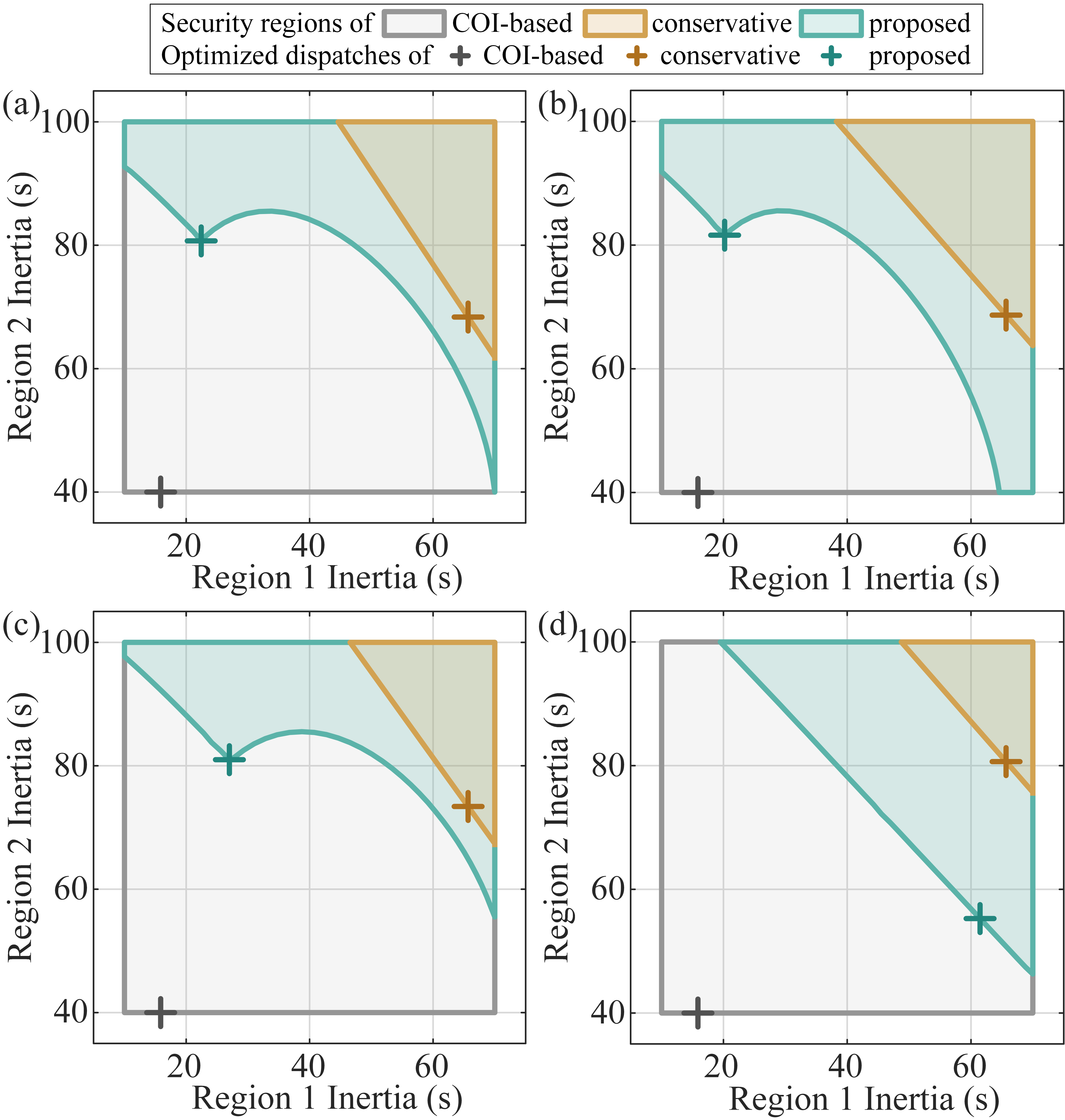} 
    \vspace{-10pt} 
	\caption{Security regions and the scheduled inertia of time periods of (a) hour 0.5 (b) hour 2, (c) hour 3, and (d) hour 4.}
    \vspace{-14pt} 
	\label{fig:CaseAnal}
\end{figure}

\begin{figure}[!t]
	\centering
	\includegraphics[width=0.45\textwidth]{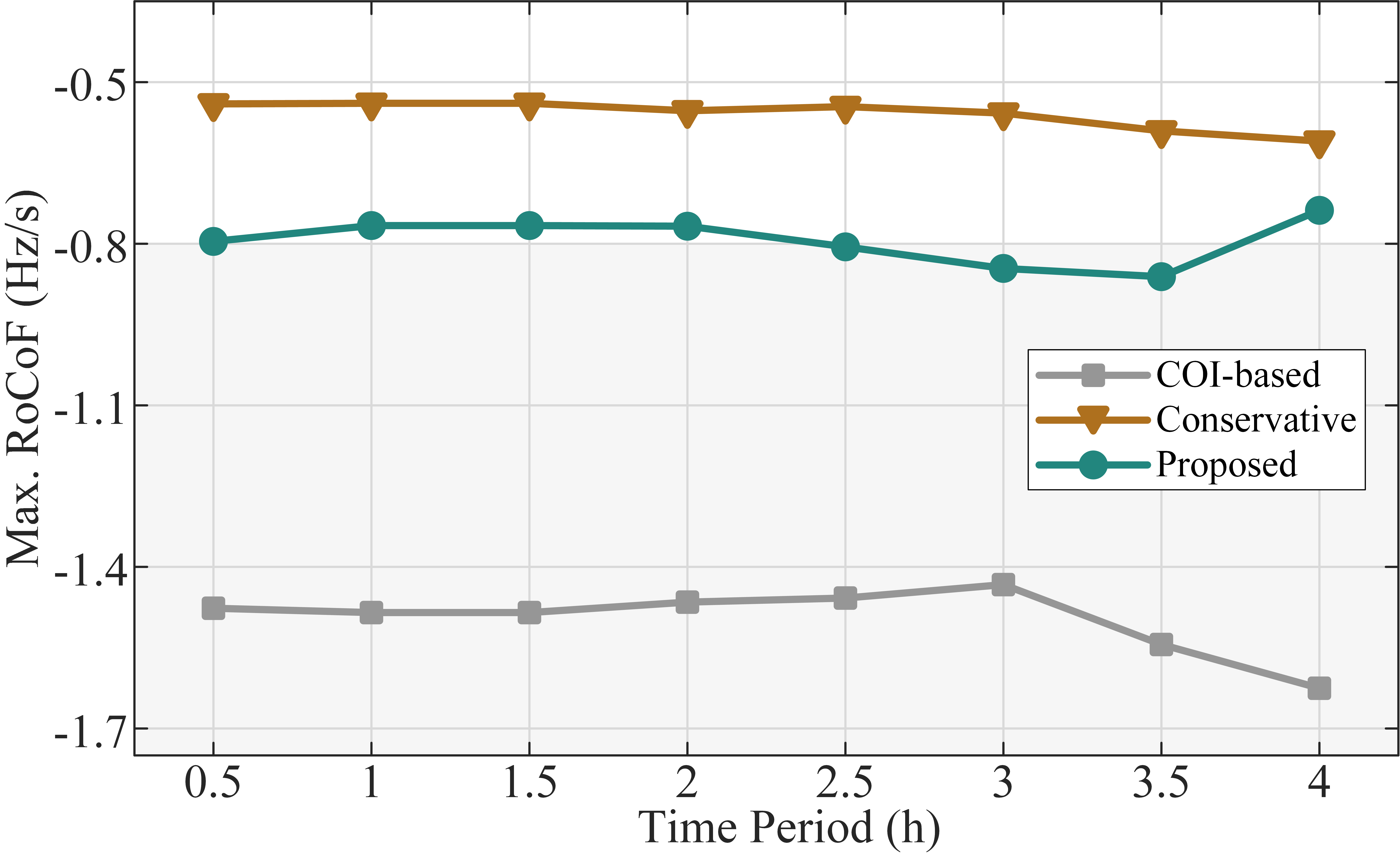} 
    \vspace{-10pt} 
	\caption{Simulated maximum RoCoF under dispatched results.}
    \vspace{-20pt} 
	\label{fig:UCRoCoF}
\end{figure}

The generation scheduling results are given in Fig. \ref{fig:UCResu}, where the fast-start SG and VI are marked. The y-axis corresponding to the inertia of regions $2$ and $3$ are compressed to give a clear view of region $1$. The three bars correspond to the COI-based, conservative, and proposed methods. Fig. \ref{fig:CaseAnal} (a)-(d) show the calculated security regions and scheduled inertia for hours $0.5$, $2$, $3$, and $4$. The real maximum RoCoF given by simulation under the dispatch result is shown in Fig. \ref{fig:UCRoCoF}. The analysis of the results is given in the following.

\subsubsection{Proposed method}

From time period hour $0.5$ to hour $2.5$, the load flow condition changes slightly. As can be verified in Fig. \ref{fig:CaseAnal}(a) and (b), the R-ISR is similar at hour $0.5$ and hour $2$. Thus, the scheduling results given in Fig. \ref{fig:UCResu} are similar from hour $0.5$ to hour $2.5$. The slight difference of regional inertia due to load flow changes is compensated by a cheap VI from IBR, i.e., G$07$, and VI from region $2$. The real maximum RoCoF shown in Fig. \ref{fig:UCRoCoF} depicts that the continuously adjustable VI can maintain the RoCoF close to the limitation without more generation start-up, which would be expensive. Starting from the time period of hour $3$, due to large variations in operating conditions, the R-ISR changes significantly as shown in Fig. \ref{fig:CaseAnal}(c) and (d). Especially at hour $4$, the RoCoF-constrained boundary changes to a straight line due to another trigonometric component in RoCoF dynamics playing a dominant role. Thus, as shown in Fig. \ref{fig:UCResu}, a fast-start SG G$02$ is committed. During hour $3.5$ and hour $4$, the VI from G$07$ is not enough and a relatively expensive VI from G$06$ starts. At the hour $4$, the inertia distribution of regions $1$ and $2$ changes due to the boundary shape in Fig. \ref{fig:CaseAnal}(d) allowing more cheap inertia from IBR in region $1$. The real RoCoF shown in Fig. \ref{fig:UCRoCoF} is near the limitation due to the flexible VI setting. Note that the RoCoF at hour $3$ and hour $3.5$ is slightly over the limit because of the error in the proposed RoCoF calculation method. This can be overcome by slightly reducing the RoCoF limitation in the UC model.

\subsubsection{COI-based method}

As can be seen in Fig. \ref{fig:CaseAnal}, the COI RoCoF-constrained boundary does not work because the COI RoCoF constraint is too loose. Thus, only a limited number of generators are committed as shown in Fig. \ref{fig:UCResu}. The real maximum RoCoF in Fig. \ref{fig:UCRoCoF} exceeds the limitation largely and will lead to frequency collapse in real-world grid.

\subsubsection{Conservative method}

Fig. \ref{fig:CaseAnal} shows that the RoCoF-constrained boundary is too tight under this method. Thus, as seen in Fig. \ref{fig:UCResu}, two fast-start SGs are turned on and VI from two IBRs are set to the maximum value, leading to high operation costs. The real RoCoF in Fig. \ref{fig:UCRoCoF} is too conservative.

\begin{table}[!t]
	\setlength\tabcolsep{3.0pt}
	\setlength{\aboverulesep}{0.0pt}
	\setlength{\belowrulesep}{1.0pt}
	\centering
	\caption{Summations of cost and time burden}
    \vspace{-8pt} 
    \begin{tabular}{cccccc}
		\toprule
		\specialrule{0em}{0.4pt}{0.4pt}
		\toprule
      \multicolumn{3}{c}{Item} & COI-based & Conservative & Proposed \\
    \midrule
    \multirow{7}[4]{*}{\makecell{Cost\\(10$^3$ \$)}} & \multirow{4}[2]{*}{Region 1} & SG & 22.45 & 22.30 & 24.12 \\
    \cmidrule{3-6}          &  & Fast-start SG & 0 & 14.73 & 3.38 \\
    \cmidrule{3-6}          &  & VI-equipped IBR & 4.89 & 48.73 & 14.79 \\
    \cmidrule{3-6}          &  & \textit{Total} & 27.35 & 85.77 & 42.29 \\
    \cmidrule{2-6}          & \multicolumn{2}{c}{Region 2} & 31.00 & 69.10 & 75.91 \\
    \cmidrule{2-6}          & \multicolumn{2}{c}{Region 3} & 36.00 & 36.00 & 36.00 \\
    \cmidrule{2-6}          & \multicolumn{2}{c}{\textit{Total}} & 94.35 & 190.87 & 154.20 \\
        \midrule
    \cmidrule{2-6}    \multirow{3}[1]{*}{\makecell{Time\\(s)}} & \multicolumn{2}{c}{Boundary calculation} & $\approx$0    & 4735  & 5176 \\
    \cmidrule{2-6}         & \multicolumn{2}{c}{MILP solving} & $\approx$0    & $\approx$0     & 1016 \\
    \cmidrule{2-6}          & \multicolumn{2}{c}{\textit{Total}} & $\approx$0    & 4735  & 6192 \\
    \bottomrule
	\specialrule{0em}{0.4pt}{0.4pt}
    \bottomrule
    \end{tabular}
	\label{tab:UCTab}
    \vspace{-20pt} 
\end{table}

Finally, the costs from different types of generators and regions are summarized in Table \ref{tab:UCTab}. The cost of the COI-based method is significantly lower than the other two methods, but the RoCoF limitation will be breached. Compared with the conservative method, the cost from fast-start SG and VI-equipped IBR in the proposed method is lower, and the costs from other parts are similar. Thus, the total cost is $80.8$\% of the conservative method, which saves costs for the regional power grid operator. The time burden is also listed in Table \ref{tab:UCTab}, where the time for frequency-constrained boundary calculation and MILP solving is listed separately. The time for the COI-based method is nearly zero. The time for calculating the frequency-constrained boundary is nearly the same for the proposed method and the conservative method, as mentioned in Section \ref{SubSec:AccuBoun}. The time for MILP solving in the proposed method is higher than that of the conservative one, which is nearly zero due to the simple linearized boundary. However, the extra time burden is not significant for hourly dispatch.

It should be noted that the time burden of the proposed method is less affected by the system scale. The boundary searching process is based on the regional COI frame, which ensures that the time burden for searching is not influenced by the size of individual regions. Similarly, the inertia security constraint, also based on the regional COI frame, is unaffected by the size of a region.

\vspace{-8pt} 
\section{Conclusion\label{SubSec:Conclu}}
\vspace{-4pt} 

This paper comprehensively studies RoCoF-constrained regional inertia security. First, the regional inertia security region is proposed. Then, the global maximum RoCoF is expressed as the maximum value of all local maxima, which are calculated at the extrema points of the trigonometric components in the RoCoF dynamics. The complete boundary of R-ISR is then expressed as the union of boundaries corresponding to these local maximum RoCoF values. A search kernel is designed to build the security boundary. Finally, the non-convex region is linearized as an inertia constraint using convex decomposition and the big-M method. The models for inertia security assessment and fast adjustment are provided.

Results on a $3$-region system show high accuracy of the proposed R-ISR boundary compared with the simulation-based method. The contour plot of maximum RoCoF shows that increasing regional inertia in one area may increase its post-disturbance RoCoF, which is counter-intuitive. The boundaries calculated by the system COI RoCoF or the conservative approach are either too loose or too tight, leading to frequency instability or high operation costs, respectively. The computation burden is mild, making it suitable for hourly dispatch.

\bibliographystyle{IEEEtran}

\vspace{-12pt} 
\bibliography{Regional_Inertia_Security}


\end{document}